\documentclass[fleqn]{aa}  
\pdfoutput=1
\usepackage{graphicx,color}
\usepackage[fleqn]{amsmath} 
\usepackage[varg]{txfonts}
\usepackage[varg]{txfonts}

\renewcommand{\div}{\ensuremath{\vec{\nabla} \cdot}}
\newcommand{\del}{\ensuremath{\dfrac{\partial}{\partial t}}}
\newcommand{\PUPs}{PUIs\ensuremath{_{H^{+}}}\xspace}
\newcommand{\PUP}{PUI\ensuremath{_{H^{+}}}\xspace}
\newcommand{\pH}{\ensuremath{_{H^{+}}}}
\newcommand{\EH}{\ensuremath{_{H^{0}}}}
\newcommand{\erf}{\ensuremath{\mathrm{erf}}}
\newcommand{\vertihori}{}
\usepackage{multicol, xspace}

\newcommand{\asr}{Adv. Space Res.}
\newcommand{\Hy}{hydrogen\xspace}
\newcommand{\He}{helium\xspace}
\newcommand{\PUI}{\ensuremath{\mathrm{PUI}}}
\newcommand{\ENA}{\ensuremath{\mathrm{ENA}}}

\newcommand{\nc}[1]{ #1}

{\end{cases}\right\rbrace}%
\begin{document}
   \title{Ionization rates in the heliosheath and in astrosheaths}

   \subtitle{Spatial dependence and dynamical relevance}

   \author{K. Scherer \inst{1} \and H. Fichtner \inst{1}
     \and H.-J. Fahr \inst{2} \and M. Bzowski \inst{3}
     \and S.E.S. Ferreira \inst{4}
          }

   \institute{Institut f\"ur Theoretische Physik IV, Ruhr-Universit\"at Bochum, 
              44780 Bochum, Germany, 
              \email{kls@tp4.rub.de, hf@tp4.rub.de}
              \and
              Argelander Institut, Universit\"at Bonn, 53121 Bonn, Germany, 
              \email{hfahr@astro.uni-bonn.de}
              \and 
     	      Polish Space Science Center, Bartycka 18A, 00-716 Warsaw, Poland, 
              \email{bzowski@cbk.waw.pl}
              \and
	      Centre for Space Research, North-West University,
              2520 Potchefstroom, South Africa
	      \email{Stefan.Ferreira@nwu.ac.za}
             }

   \date{Received; accepted}

 
  \abstract
  {In the heliosphere, especially in the inner heliosheath, mass-,
    momentum-, and energy loading induced by the ionization of neutral
    interstellar species plays an important, but for some species,
    especially \He, an underestimated role.}
   {We discuss the implementation of charge exchange and electron impact processes for
     interstellar neutral \Hy and \He and their implications 
     for further modeling. Especially, we emphasize the importance of
     electron impact and a more sophisticated numerical treatment of
     the charge exchange reactions. Moreover, we discuss the
     non-resonant charge exchange effects.}
   {The rate coefficients are discussed and the influence of the
     cross-sections in the (M)HD equations for
     different reactions are revised as well as their representation
     in the collision integrals.}
   {Electron impact is in some regions of the heliosphere, particularly
     in the heliotail, more effective than charge exchange, and the
     ionization of neutral interstellar \He contributes about 40\%
     to the mass- and momentum loading in the inner heliosheath. The
     charge exchange cross-sections need to be modeled with higher
     accuracy, especially in view of the latest developments in their
     description.}
   {The ionization of \He and electron impact
     ionization of \Hy needs to be taken into account for the
     modeling of the heliosheath and, in general, astrosheaths. Moreover,
     the charge exchange cross-sections need to be handled in a more
     sophisticated way, either by developing better analytic
       approximations or by solving the collision integrals
       numerically. }


   \maketitle
%

\section{Introduction}
\subsection{General Aspects}

It is well-known that not only the interstellar plasma but also the
interstellar neutral gas is influencing the large-scale structure of
the heliosphere
\citep{Baranov-Malama-1993,Scherer-Ferreira-2005a,Mueller-etal-2008,Zank-etal-2013}. This
influence is a consequence of the coupling of the neutral gas
(consisting mainly of \Hy and \He) to the solar wind plasma (mainly
protons with a small contribution of $\alpha$ particles (He$^{2+}$))
via the charge exchange, photo-ionization, and electron impact
processes. Elastic collisions and Coulomb scattering are discussed
in \citet{Williams-etal-1997}. 

In the self-consistent models of heliospheric dynamics, so far, only
the influence of neutral \Hy is considered by taking into account its
charge exchange with solar wind protons and its ionization by the
solar radiation
\citep[e.g.\
][and references therein]{Fahr-Rucinski-2001,Pogorelov-etal-2009b,
  Alouani-Bibi-etal-2011},
The dynamical relevance of both the electron impact ionization of \Hy,
although recognized by \citet{Malama-etal-2006}, and the
photo-ionization of \He, although recognized as being filtrated in the
inner heliosheath \citep{Rucinski-Fahr-1989,Cummings-etal-2002},
have not yet been explored in detail. There is only one attempt to
include \He self-consistently in the heliospheric modeling,
namely \citet{Malama-etal-2006}, in which the emphasis is on the
additional ram pressure due to the charged \He ions.

In recent years the discussion has rather concentrated on the correct
cross-section for the charge exchange between a proton and a \Hy
atom. While it has been demonstrated by \citet{Williams-etal-1997}
that the modeling results regarding the large-scale structure
(location of the termination shock and heliopause) are insensitive to
use of the alternative cross-sections given by \citet{Fite-etal-1962}
and \citet{Maher-Tinsley-1977} it was revealed that it is of importance
for the neutral gas (shape of the \Hy wall and densities inside
the termination shock
see, e.g. \citet{Baranov-etal-1998} and \citet{Heerikhuisen-etal-2006}.
Subsequently, the significance of the revision of these cross-sections
by \citet{Lindsay-Stebbings-2005} has first been recognized by
\citet{Fahr-etal-2007} and then been
discussed further in the context of global heliospheric modeling by
\citet{Mueller-etal-2008}, \citet{Bzowski-etal-2008}, \citet{Izmodenov-etal-2008}.

It has also been recognized that electron impact ionization is not
only of importance for the neutral gas distribution
\citep[e.g.,][]{Rucinski-Fahr-1989,Moebius-etal-2004, Izmodenov-2007}
but, to some extent, also for the large-scale structure of the
heliosphere
\citep{Fahr-etal-2000, Scherer-Ferreira-2005a, Malama-etal-2006}.
These studies, however, contain neither comparison nor analysis of the
electron impact ionization rates to those of the other two processes
nor a systematic analysis of their dynamical influence.

Besides addressing the first of these issues in the present paper, we
will show that not only the ionization of neutral \Hy influences
the dynamics in the heliosphere, but that also neutral \He must be expected to
play a role. Furthermore, we discuss charge exchange reactions of
\Hy and \He in view of their relevance to astrospheres,
particularly to astrosheaths. In astrospheres the relative speed between a
stellar wind and the flow of neutrals from the interstellar medium
can be up to an order of magnitude higher than in the solar case, see
 for example the M-dwarf V-Peg 374
\citep{Vidotto-etal-2011} or for hot stars  \citep[e.g.\ ][]{Arthur-2012}. 

Before quantitatively addressing these topics, we briefly indicate
further heliospheric and astrophysical applications for which the
ionization rates studied here are of interest, too.

The charge exchange process has recently gained more interest as it is
responsible for the outer-heliospheric production of energetic neutral
atoms \citep[see][ for a comprehensive review]{Fahr-etal-2007} that are
presently observed with the IBEX mission,
\citep{McComas-etal-2009a,Funsten-etal-2009,Dayeh-etal-2012,McComas-etal-2012}. In this
context, \citet{Grzedzielski-etal-2010} investigated the distribution
of different pickup ion (PUI) species, being products of ionization
processes in the inner heliosheath and \citet{Borovikov-etal-2011}
discussed their influence on the plasma state of the latter.
Furthermore, \citet{Aleksashov-etal-2004} started to explore the
influence of charge exchange of \Hy atoms with the solar wind
protons and their impact on the structure of the heliotail. The charge exchange
processes of heavier elements are studied in connection with the X-ray
production in the inner heliosphere \citep{Koutroumpa-etal-2009},
while these processes in the LISM are discussed by \citet{Provornikova-etal-2012}.


Astrophysical scenarios for which the use of correct ionization rates
is crucial \citep{Nekrasov-2012}, comprise the X-ray emission from galaxies
\citep{Wang-Liu-2012}, from the interstellar medium
\citep{Avillez-Breitschwerdt-2012} and from hot stars
\citep{Pollock-2012}, the influence of neutral atoms on astrophysical
shocks \citep{Blasi-etal-2012, Ohira-2012}, and the jets of active
galactic nuclei \citep{Gerbig-Schlickeiser-2007}.


The present study concentrates on the relative importance of the
different ionization processes for the heliosheath and for
astrosheaths, i.e.\ the regions between the solar/stellar wind
termination shock/s and the helio-/astropauses, in order to prepare
their later incorporation into corresponding self-consistent (M)HD
modeling with the BoPo- \citep{Scherer-Ferreira-2005a} and the
CRONOS-code
\citep[see][for an application and the references
therein]{Wiengarten-etal-2013}. For the computations here, with which
we demonstrate the significance and dynamical relevance of electron
impact ionization of \Hy and of \He in the entire heliosphere, we
employ the well-established heliospheric model by
\citet{Fahr-etal-2000} and \citet{Scherer-Ferreira-2005a} where charge
exchange between neutral and ionized \Hy, electron impact of \Hy, and
photo-ionization of \Hy is included. The consideration of elastic
  collisions and Coulomb scattering \citep{Williams-etal-1997} are
  beyond the scope of this paper.

In section~\ref{sec:2} we first give a short introduction to the
BoPo-code and continue with a  discussion of the charge
exchange and electron impact cross-sections in section~\ref{sec:3}. In
section~\ref{sec:4} we discuss the interaction terms used in the
Euler equations (see appendix~A). Then, in section~\ref{sec:5} we concentrate
on the electron impact and its contribution to the
interaction terms, which is then followed by a discussion of the
neutral interstellar \He loss in the inner heliosheath in
section~\ref{sec:6}. Section ~\ref{sec:cx} contains a discussion of
the non-resonant charge exchange and its relevance to the outer heliosheath. Finally we
assemble our ideas and critically assess our findings in the concluding
section~\ref{sec:7}.

\subsection{Astrospheres and the heliosphere}

An astrosphere is a generalization of the heliosphere. The physics is
the similar, except that for huge astrospheres, like those around hot
stars \citep{Arthur-2012}, where a cooling function needs to be
included beyond the termination shock. Additionally, for hot stars the
Str\"omgren sphere can be larger then the astrosphere, and thus the
photo-ionization can be dominant in the entire astrosphere.
 Nonetheless, around hot stars can exits HI regions
  \citep{Arnal-2001,Cichowolski-etal-2003} and, thus, ionization
  processes become important. Therefore, a better concept than the
  classical Str\"omgren sphere is that of an HII-region with a
  variable degree of ionization. For example, for the Sun the
  Str\"omgren radius $R_{S}=2.2\cdot10^{10}$\,cm \citep{Fahr-1968} is
  smaller than the solar radius, while  its HII region is
  considerably larger \citep{Lenchek-1964,Ritzerveld-2005}.

  There exist cool stars like V-Peg 374 \citep{Vidotto-etal-2011}
  which have stellar wind speeds ($\approx 2000$\,km/s) much higher
  than that of the solar wind. Also for nearby stars the relative
  speeds between the star and the interstellar medium can differ by a
  factors between 0.2 to 2 compared to that surrounding the solar
  system \citep{Wood-etal-2007}. For these high relative speeds
  cross-sections of other processes and different species may become
  important.

  In most of the stellar wind models
  \citep[e.g.\ ][]{Lamers-Cassinelli-1999} the wind passes through one
  or more critical points (surfaces) after which it freely
  expands. Such a freely expanding continuous blowing wind is
  physically similar to the solar wind. In an ideal scenario without
  neutrals and cooling the solutions can simply be scaled from one
  scenario into the other.  

The models for astrospheres are based on MHD equations (see
appendix~A), in the same way as for the heliosphere. Because a fleet
of spacecraft has been or is exploring the heliosphere by in-situ as
well as remote measurements, the knowledge of that special
astrosphere is much more detailed than that of astrospheres. In the
following we do not make a difference between astrospheres or the
heliosphere. All what is discussed applies as well to the heliosphere
as similarly to astrospheres immersed in partly ionized clouds of
interstellar matter.

One interesting feature of ome nearby astrospheres is their
hydrogen walls which are built beyond the astropauses by charge
exchange between interstellar hydrogen and protons. The feature can be
observed in Lyman-$\alpha$ absorption \citep{Wood-etal-2007}, which in
turn allows to determine the stellar wind and interstellar parameters
at some nearby stars. Because the hydrogen wall is built up in
the interstellar medium, where the temperatures are low ($<10^{4}$\,K)
and in case of the heliosphere increase only by approximately a
  factor two towards the heliopause
the charge exchange process involved is that between protons and \Hy
as well as some \He reactions, like He$^{+}$+He, He$^{2+}$+He and
  He$^{+}$+He$^{+}$,
which have large cross-sections even at low energies. Due to the low
temperatures electron impact is not effective, because the involved
energies are less than the ionization energy.  Up to now, only a \Hy
wall was observed, which, nevertheless, allows for some insights into
the structure of nearby astrospheres.  An \He wall as a result of
helium-proton charge exchange was found not to exist
\citep{Mueller-Zank-2004b}. Nevertheless, some helium-helium reactions
with sufficiently large cross-sections (see Fig.~\ref{Fig2} or
tables~\ref{tab:1} and~\ref{tbb}) may result in a helium wall.

\section{The heliosphere model\label{sec:2} }

\begin{figure}
   \centering
   \includegraphics[width=0.95\columnwidth]{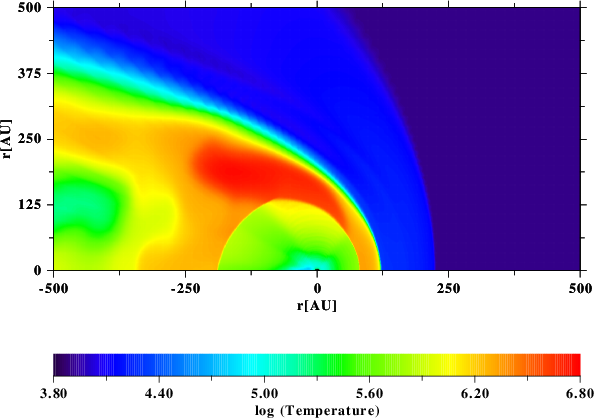}
   \caption{The proton temperature distribution in the dynamic heliosphere
     \citep{Scherer-Ferreira-2005a}. Over the solar pole the features
     of a high-speed
     stream can be identified. Due to the influence of the \PUPs the
     temperature inside the termination shock is a few
     $10^{5}$\,K, corresponding to thermal energies up to a few tens
     of eV. The temperature in the inner heliosheath is $10^{6}$\,K, which
     is inferred by IBEX
     \citep{Livadiotis-etal-2011}.}
              \label{Fig1}%
    \end{figure}

\begin{figure*}
  \includegraphics[width=1.0\columnwidth]{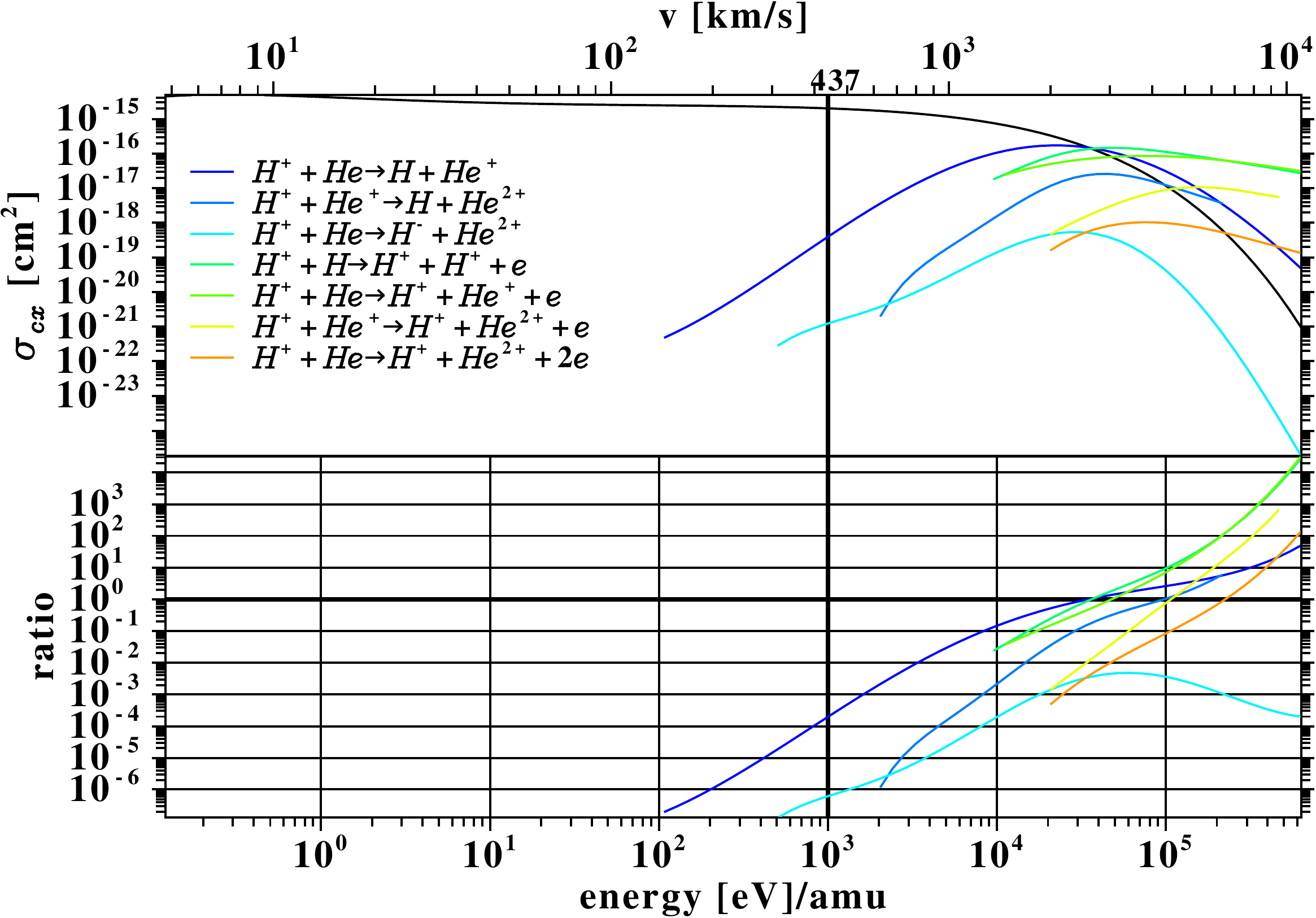}\vertihori
  \includegraphics[width=1.0\columnwidth]{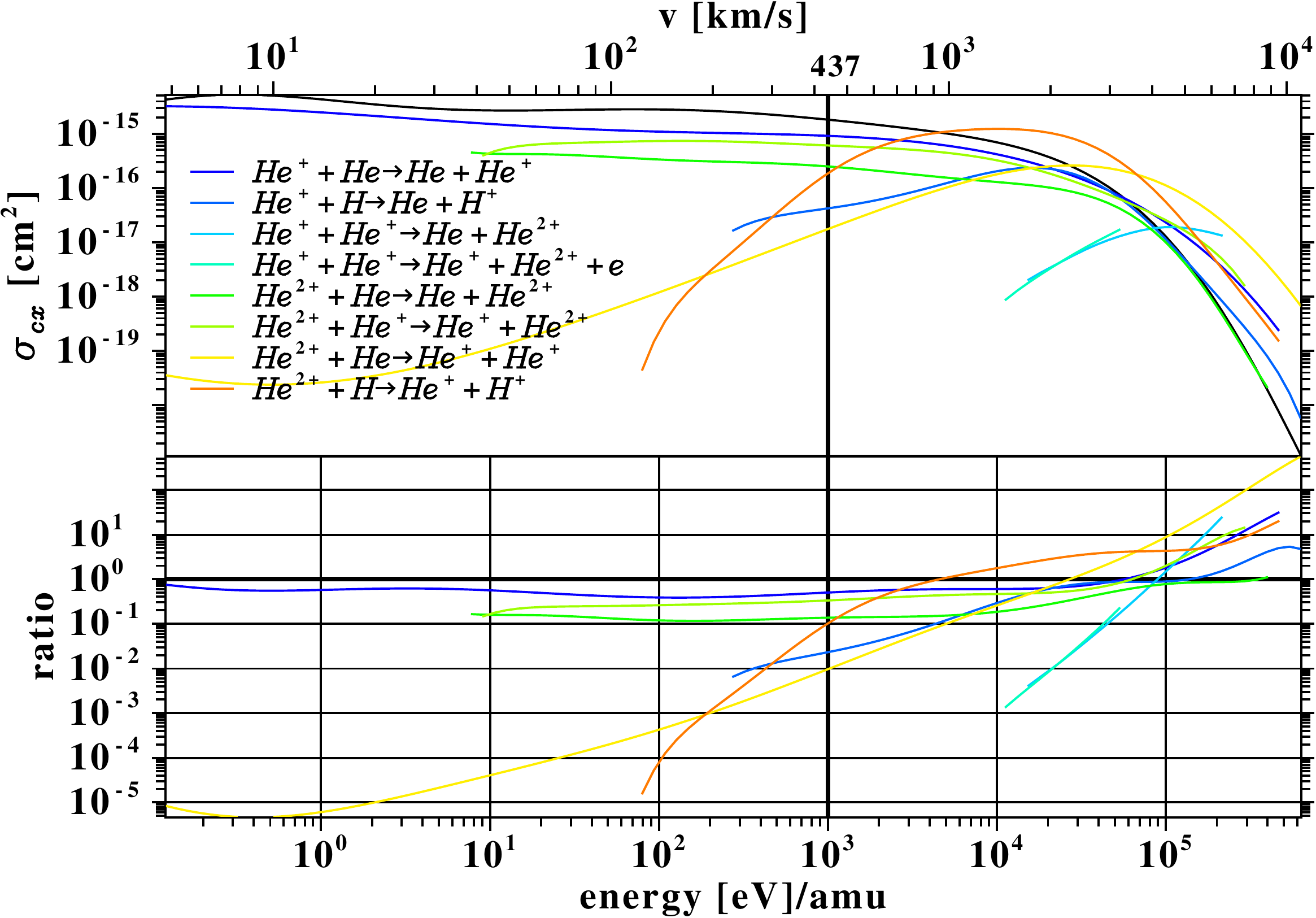}
  \caption{The charge exchange cross-section as function of energy per
    nucleon for protons (left panel) as well as $He^{+}$-ions and
    $\alpha$-particles (right panel) of the solar wind with
    interstellar \He and \Hy. In the upper part of both panels the
    cross-sections are shown, while the lower parts show the ratio to
    $\sigma_{cx}(H^{+}+H\rightarrow H+H^{+})$. The black curve in both
    panels is the reaction $H+H^{+}\rightarrow H^{+}+H$. As can be
    seen in the lower panel the reactions He$^{+}$+He,
    He$^{2+}$+He, and He$^{2+}$+He$^{+}$ have similar cross-sections
    than that of H+p, and thus are important in modeling the dynamics
    of the large-scale astrospheric structures. Note the different y-axis
  scales between different panels.}
  \label{Fig2}%
\end{figure*}

For this analysis a detailed model of the heliosphere is not
necessarily needed. Nevertheless, we will use a snapshot of a dynamic
heliosphere model computed with the BoPo-model code
\citep{Scherer-Ferreira-2005a}. This model includes three species,
namely, protons, neutral \Hy and, from the latter, newly created ions,
the pickup protons (\PUPs). The effective temperature distribution,
i.e.\ the weighted sum of proton and PUI temperatures of the model is shown in
Fig.~\ref{Fig1}. The modeled effective temperature reproduces nicely
those inferred by IBEX \citep{Livadiotis-etal-2011}.

\begin{figure*}
  \includegraphics[width=1.0\columnwidth]{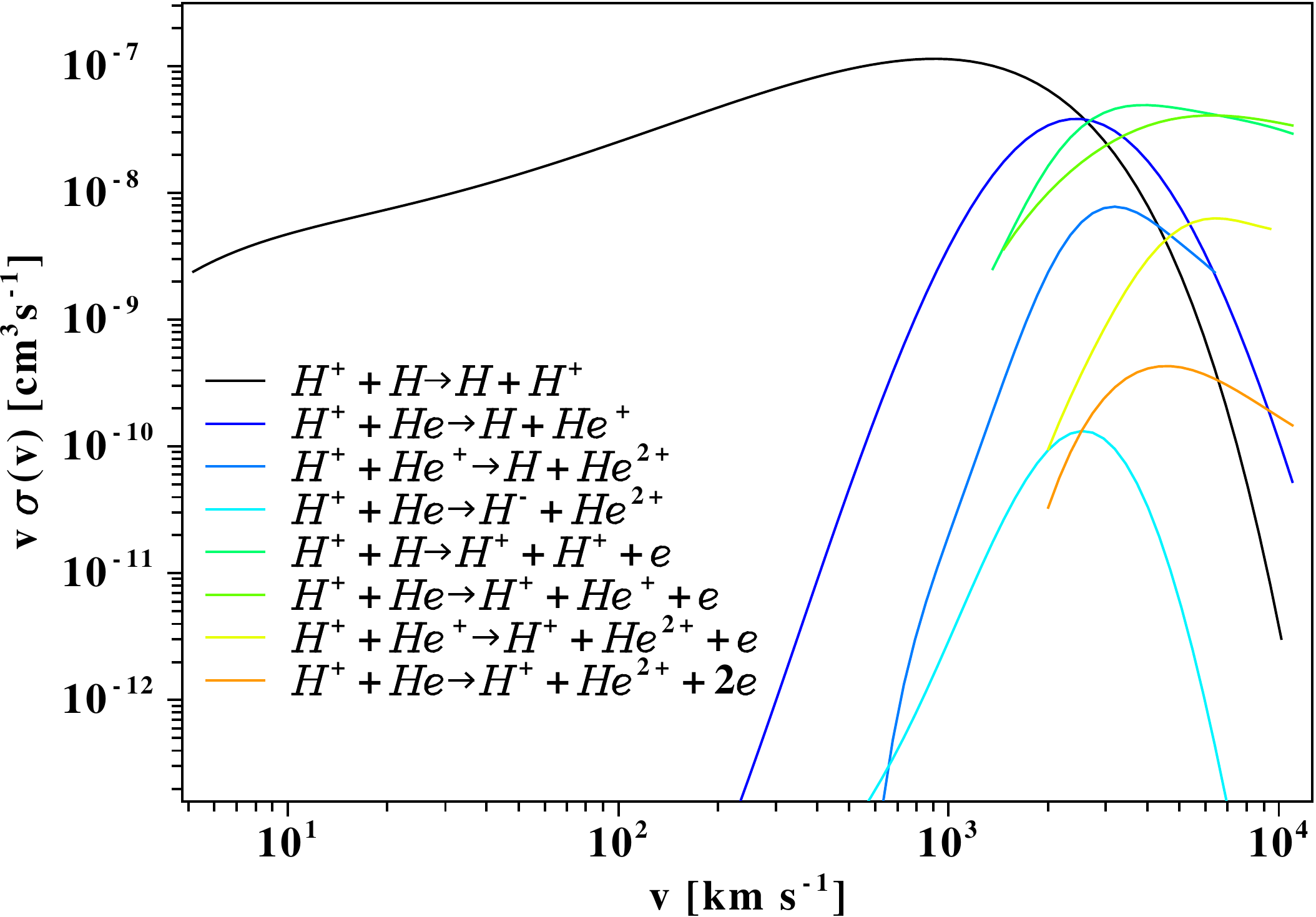}\vertihori
  \includegraphics[width=1.0\columnwidth]{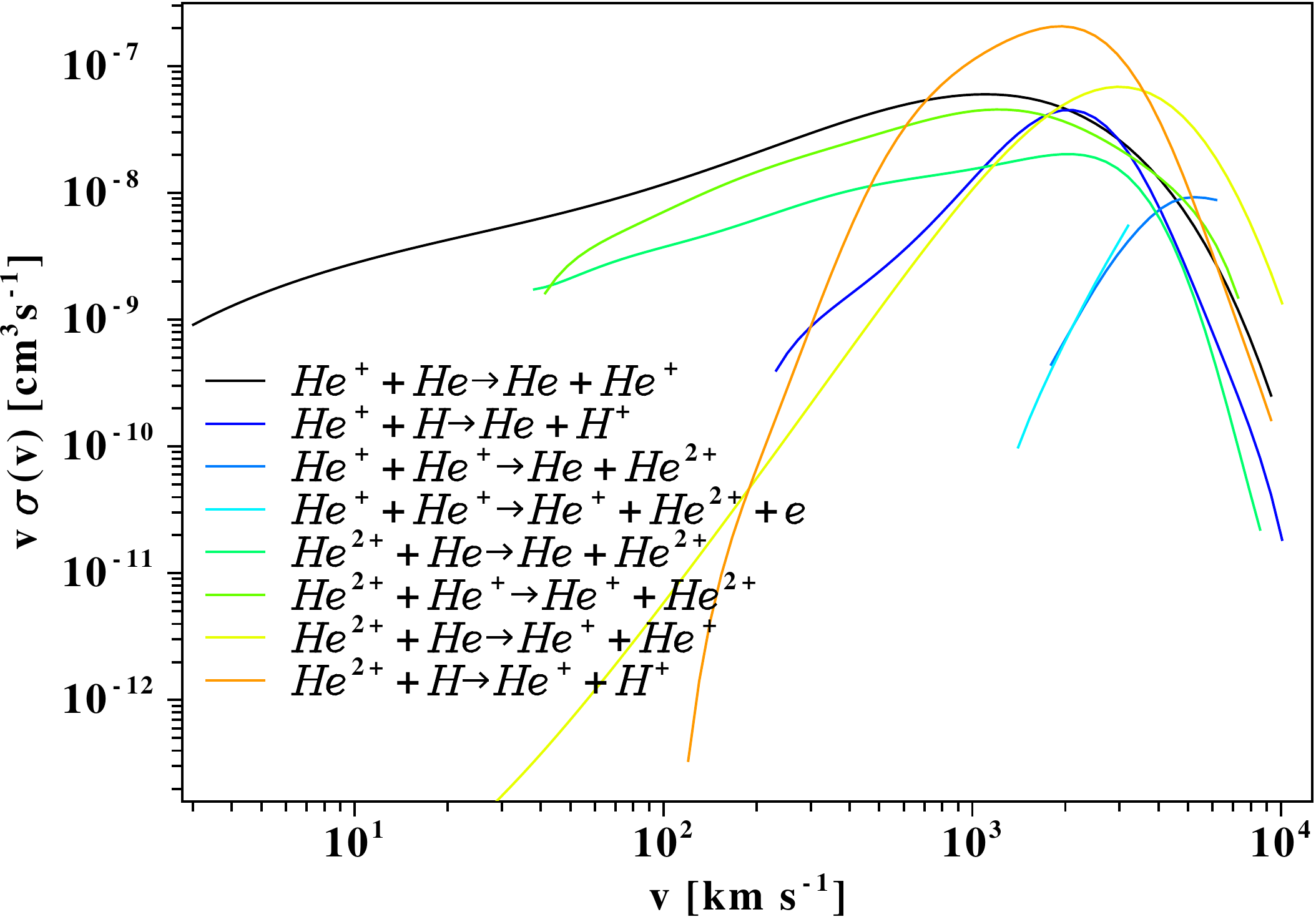}
\caption{Shown are the rate coefficients $\beta(v)=v\sigma(v)$ instead of the
  cross-sections as in Fig.~\ref{Fig2}. We assumed that the there is no
    difference in the relative speeds from the collision integrals
    $v_{rel,ij}^{r,I_{ coll}}$ and those to be used for the
    cross-sections $v_{rel,ij}^{r,\sigma}$, i.e.\ $v =
    v_{rel,ij}^{r,I_{ coll}} = v_{rel,ij}^{r,\sigma}$  \label{Fig2b}}
\end{figure*}

The \PUPs are created by a resonant charge exchange process between a neutral
\Hy atom and a proton:
\begin{equation}
  H^{+}+ H \rightarrow H + H^{+}
\end{equation}
in which an electron is exchanged between the reaction
partners. Because the first reactant is a fast proton, the neutral \Hy
atom resulting from this interaction is an energetic (fast) neutral
atom (ENA), which here is assumed to leave the heliosphere without
further interaction. The second reactant, a slow interstellar \Hy atom,
becomes ionized and is immediately picked up by the electromotive
force of the heliospheric magnetic field frozen-in the solar wind. The
initial \PUP velocity distribution is  ring-like with a
maximum of twice the solar wind speed
\citep{Vasyliunas-Siscoe-1976,Isenberg-1995,Gloeckler-Geiss-1998}.
It very quickly becomes pitch-angle isotropized, thus transforming
into a nearly spherical shell distribution. 

During this charge exchange process the total ion density does not
change, but, because of the velocity difference between the proton and
neutral \Hy atom, the solar wind momentum is altered (momentum
loading) as well as the energy of the fluid, i.e.\ the temperature of
the solar wind increases and momentum decreases, because it is lost
due to the escaping ENA \citep{Fahr-Rucinski-1999}. This temperature
increase was observed by the Voyager spacecraft
\citep{Richardson-Wang-2010}, while model runs including only a single
proton fluid resulted in temperatures of a few hundred Kelvin at the
termination shock
\citep[for an example see Fig.1 in][]{Fahr-etal-2000}. This contrasts
with the above mentioned multifluid and multispecies models with
$10^{5}$\,K at the same location (see Fig.~\ref{Fig1}).  The momentum
loading leads to a slowdown of the solar wind, which is indeed
observed \citep{Richardson-Wang-2010}, in agreement with model
attempts
\citep{Fahr-Fichtner-1995,Fahr-Rucinski-2001,Fahr-2007,Pogorelov-etal-2004,Alouani-Bibi-etal-2011}.
Moreover, including neutrals removes the Mach-disks
\citep{Pauls-etal-1995,Mueller-etal-2001} and thus change the large
scale structure of the astro-/heliosphere and, thus, again
demonstrates the importance of including charge exchange processes for
an adequate description.

In the dynamic BoPo-model \citep{Fahr-etal-2000,Scherer-Ferreira-2005a} also the
electron impact:
\begin{equation}
H+e^{-} \rightarrow H^{+}+ 2 e^{-}
\end{equation}
in the inner heliosheath, i.e.\ the region between the termination
shock and the heliopause is included.

Additionally, the photo-ionization inside the termination shock is
included.  This leads not only to the above described momentum- and
energy-loading but also to a mass loading, because new ions are
generated.

We briefly mention a few complications which are usually not taken
into account. As reported by
\citet{Lallement-etal-2005,Lallement-etal-2010} the direction of the
 deflected \Hy and \He inflow can differ by $4^{o}$, see however
the IBEX results discussed in \citet{Moebius-etal-2012}. This additional
complication was only modeled by \citet{Izmodenov-Baranov-2006},
\nc{while a lot of effort was put in the modeling of the hydrogen deflection
  plane \citep{Pogorelov-etal-2009b,Opher-etal-2009, Ratkiewicz-Grygorczuk-2008}}. The
latest IBEX observations \citep{Saul-etal-2013} show that the hydrogen
peak is moving in longitude during the solar cycle, which according to
these authors is caused by the changing radiation pressure close to
the Sun, which affects the effective gravitational force. In the outer
heliosphere this effect is assumed to be negligible and, hence, is
independent from charge exchange and electron impact processes, it is
not taken into account further on.

Since decades \citep{Fahr-1979} it has been known, that close to the
Sun the \He is focused in the downwind direction, while the \Hy is
defocused, again the latter is caused by the interaction between the
radiation pressure and gravitation acting on these particles. For a
more detailed analysis of the trajectories see the recent paper by
\citet{Mueller-2012}.

Because we are mainly interested in the large-scale structure far away
from the Sun, we can neglect this additional complication, for which
the \Hy and \He fluids have to be treated kinetically \citep[for \Hy
see ][]{Osterbart-Fahr-1992}. 

The ionization processes we discuss are independent from the
underlying (M)HD model. If the neutral fluid is treated kinetically
\citep{Izmodenov-2007,Heerikhuisen-etal-2008} the distribution
functions of the collision integrals cannot be handled by two
Maxwellians, but one is determined by the solution of the kinetic
equation. Thus the details of the collision terms may vary, but the
following principal discussion remains true. Moreover, to avoid the
solution of kinetic equations, a multifluid approach is often used
\citep{Heerikhuisen-etal-2008, Alouani-Bibi-etal-2011,
  Prested-etal-2012} with several fluids in different regions.

Therefore, in what follows we use the BoPo-code
\citep{Scherer-Ferreira-2005a} results to visualize and emphasize the
aspects discussed (other (M)HD models would be equally suitable), with
intention to draw the attention to those aspects and to point out the
need of a self-consistent model including them.

In the next section we discuss some additional charge exchange
processes with \Hy and \He as well as electron impact of
\He.

%

\section{Charge exchange and electron impact cross-sections\label{sec:3} }

%
Very detailed analysis of the ionization processes at 1\,AU taking
into account also temporal variations due to the solar cycle can be
found in
\citet{Rucinski-etal-1996,Rucinski-etal-1998,Rucinski-etal-2003},
\citet{Bzowski-etal-2013} and \citet{Sokol-etal-2012}. This is much
more complicated for the outer heliosphere, especially in the inner
heliosheath, where temporal and spatial variations are mixed and
cannot be disentangled, due to the subsonic character of the fluid. For
our purpose, to demonstrate the importance of the different effects,
it is sufficient to use a simplified approach, i.e.\ assuming a
stationary model, and discussing the effects along a line of sight
between the termination shock and the heliopause in the nose region.

\subsection{Charge exchange}
In Fig.~\ref{Fig2} we show the charge exchange cross-sections
$\sigma_{cx}$ as functions of energy per nucleon between protons (left
panel) as well as $He^{+}$-ions and $\alpha$-particles (right
panel) with \Hy as well as \He. In the lower part of the panels the
different cross-sections are normalized to that of the above mentioned
``standard'' reaction $H^{+}+H\rightarrow H+H^{+}$.  All
cross-sections discussed here are taken from the ``Redbook''
\mbox{http://www-cfadc.phy.ornl.gov/redbooks/redbooks.html} and can be
accessed via the ``ALADDIN'' webpage
(\mbox{http://www-amdis.iaea.org/ALADDIN/collision.html}). While for
the discussion below the exact values of the cross-sections are not
important, we still refer the reader to
\citet{Arnaud-Rothenflug-1985}, \citet{Badnell-2006},
\citet{Cabrera-Trujillo-2010}, \citet{Kingdon-Ferland-1996},
\citet{Lindsay-Stebbings-2005} for more information.  Nevertheless, if
the cross-sections are required in modeling efforts the more modern
results should be taken into account. For the convenience of the
reader the rate coefficients $\beta(v)=v \sigma(v)$ are presented in
Fig.~\ref{Fig2b}.

From the left panel of Fig.~\ref{Fig2} one can see that
$\sigma_{cx}(H^{+}+H\rightarrow H+H^{+})$ is roughly in the range of
$10^{-15}$\,cm$^{2}$ below 1\,keV, i.e.\ the range of interest for
heliospheric models.  All other cross-sections $\sigma_{cx}$ between
protons and neutral H or He are orders of magnitude smaller for slow
solar/stellar wind conditions.  In the high-speed streams and
especially in coronal mass ejections the cross-sections like
$\sigma_{cx}(H^{+}+He\rightarrow H+He^{+})$, 
$\sigma_{cx}(H^{+}+H\rightarrow H^{+}+H^{+}+e$) and
$\sigma_{cx}(H^{+}+He\rightarrow H^{+}+He^{+}+e$) can become of the
same magnitude as $\sigma_{cx}(H^{+}+H\rightarrow H+H^{+})$.  For
astrospheres with stellar wind speeds in the order of a few thousand
km/s the energy range is shifted toward 10\,keV up to 100\,keV and
other interactions, like non-resonant charge-exchange processes, need
to be taken into account.

While the cross-section $\sigma_{cx}$ between $\alpha$-particles and
neutral H or He compared to the
$\sigma_{cx}(H^{+}+H\rightarrow H+H^{+})$ reaction seem not to be
negligible in and above the keV-range (Fig.~\ref{Fig2}, right
panel), the solar abundance of $\alpha$-particles is only 4\% of that
of the protons, so that the effect seems to be small. Nevertheless,
the mass of He or its ions is roughly four times that of (charged) H,
and thus may play a role in mass-, momentum-, and energy loading.

 Most of the cross-sections displayed in Fig.~\ref{Fig2} are not
  suitable for use in a numerical code, because the existing data are
  fitted in a limited energy range by Chebyshev polynomials as taken
  from the ``Aladdin'' webpage. Using these approximations outside
  that range leads to unreliable results. For future simulations these
  cross-sections need to be extrapolated to the whole energy range
  required for the heliosphere and astrospheres, i.e.\ from energies
  about 1\,eV to a few 100\,keV.

  From Tables~\ref{tab:1} and~\ref{tbb} one sees that the reactions
  H+He$^{+}\rightarrow$p+He and H+He$^{2+}\rightarrow$p+He$^{+}$ are a
  few percent of the H+p-reaction, but because \He is involved, one
  may get a change in the total density of the governing equations in
  the ten percent range. The depletion of \He by charge exchange
  was discussed by \citet{Mueller-Zank-2004c}, who concluded that it
  can amount to 2\%.

  According to \citet{Slavin-Frisch-2008} the ionization fraction in
  the interstellar medium of He$^{+}$ and He is about 0.4, and thus
  the He$^{+}$ abundances are about 2/3 that of He. Form
  Fig.\ref{Fig2} and the Tables~\ref{tab:1} and~\ref{tbb} one easily
  can see that the He$^{+}$+He reaction has a similar cross-section
  than that of p+H. The He$^{+}$ fluid should follow a similar pattern
  than the p-fluid in front of the heliopause: The \He ions are slowed
  down and, via charge exchange with the neutrals, must be expected
  build a \Hy wall, respectively a \He wall. The additional reaction
  He$^{+}$+He$^{+}\rightarrow$He+He$^{2+}$ with similar
  cross-sections can support the creation of the \He wall.  According
  to \citet{Mueller-2012} 10\% of the neutral \He will become ionized
  in front of the heliopause. To determine the \He wall a
  self-consistent model is needed, which includes the three species
  He, He$^{+}$, and He$^{2+}$ and all their reacting channels.

  Another consequence of the high He$^{+}$ abundances as modeled by
  \citep{Slavin-Frisch-2008} is that the total ion mass density is
  larger by a factor than the proton mass density. The latter is
  usually used to estimate the Alfv\'en
  speed, which is proportional to 1/$\sqrt{\rho}$, in the interstellar
  medium, which will be reduced by 10\% taking into account the total
  ion mass, i.e.\ the additional He$^{+}$ abundances.

  For astrospheres with higher stellar wind speeds other reactions
  like p+He$^{+}\rightarrow$H+He$^{+2}$, become important too.
  A brief discussion is given in section~\ref{sheath}.

\subsection{Electron impact}
\begin{figure}
  \centering
  \includegraphics[width=1.0\columnwidth]{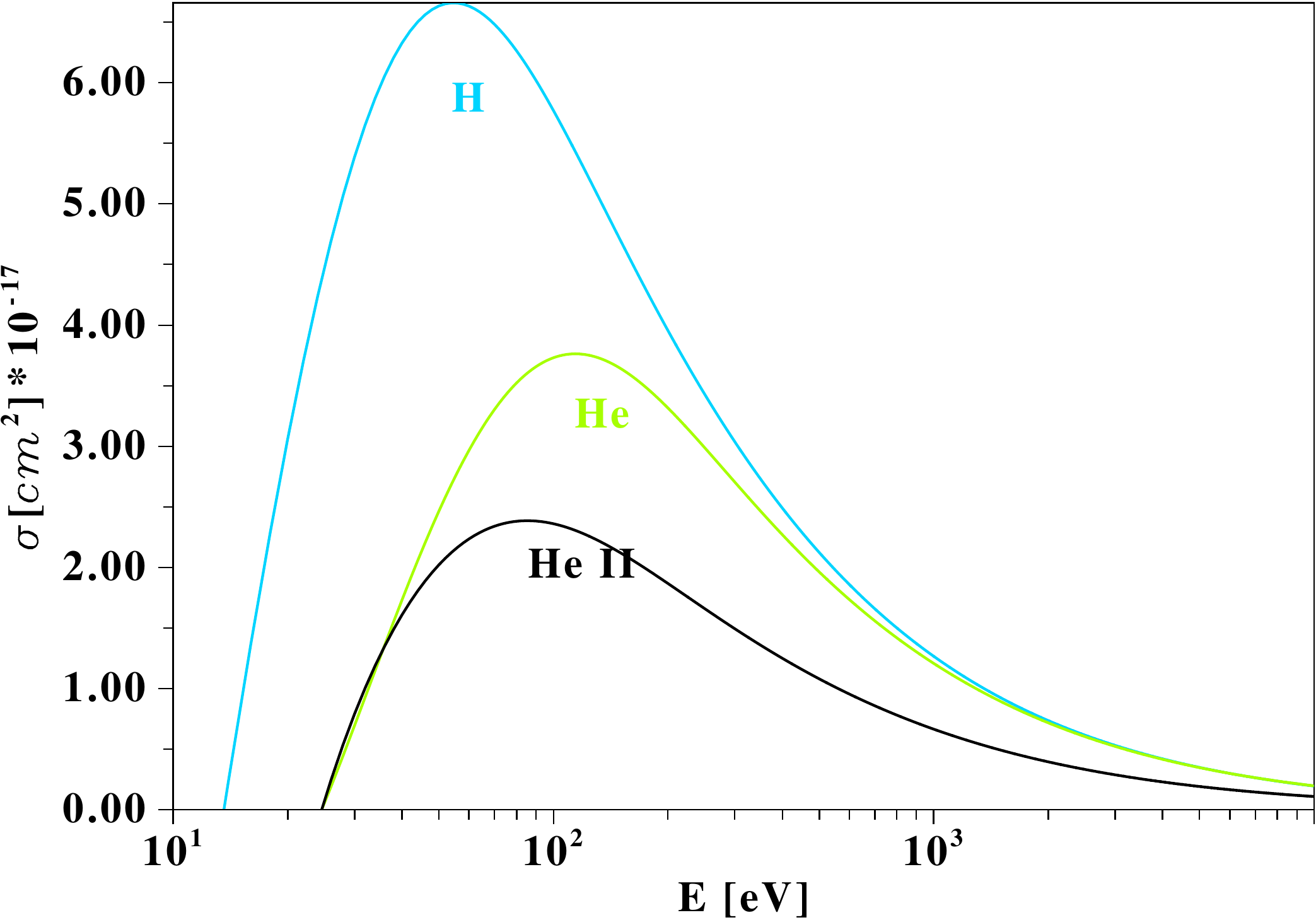}
  \caption{The electron impact cross-section $\sigma^{ei}$  in units of
    $10^{-17}$\,cm$^{2}$ per particle, for the three reactions:
    H+e$\rightarrow$H$^{+}$+2e, He+e$\rightarrow$He$^{+}$+2e, He+e$\rightarrow$He$^{2+}$+3e. }
  \label{Fig3}%
\end{figure}

The electron impact cross-section $\sigma^{ei}$ is shown in
Fig.~\ref{Fig3}. The thermal speed of the electrons is calculated
assuming thermal equilibrium between protons and electrons
\citep{Malama-etal-2006, Moebius-etal-2012}. This is done, 
because the electron temperature has, so far, neither been modeled nor
observed in the outer heliosphere, while for the inner heliosphere
(upstream of the termination shock) it has been modeled
\citep{Usmanov-Goldstein-2006} and a limited set of observations
  in the heliocentric distance range between 0.3 to about 5\,AU is
  available \citep[e.g.\ ][]{Maksimovic-etal-1997}. Additionally, care must be taken,
because the electron temperature most probably increases during the passage of
the plasma over the termination shock
\citep{Fahr-etal-2012}. Nevertheless, we follow the assumption by
\citet{Malama-etal-2006}, according to which the electrons are in thermal equilibrium
with the proton plasma (including \PUPs). The
cross-sections are calculated after \citet{Lotz-1967a} and
\citet{Lotz-1970}

\begin{eqnarray}
\sigma_{E_{rel}} &=& \sum\limits^{N}_{i=1} a_{i} q_{i}
\dfrac{\ln(E_{rel}/K_{i})}{E_{rel}K_{i}}\\\nonumber
&&\left(1 - b_{i} \exp\left[-c_{i}
    \left(\frac{E_{rel}}{K_{i}}-1\right)\right]\right);
\quad E_{rel}\ge K_{i} \label{eq:1}
\end{eqnarray}
where the index $i$ runs over all relevant subshells up to $N$. For
example for \Hy-like atoms $N=1$, for \He $N=2$, see
\citet{Lotz-1970}.  $K_{i}$ is the ionization energy in the i$^{th}$
subshell. The coefficients $a_{i}, b_{i}, c_{i}, q_{i}$ are tabulated
in \citet{Lotz-1970}.

Also other approaches exist to describe the electron impact
cross-sections
\citep{Mattioli-etal-2007,Mazzotta-etal-1998,Voronov-1997}. They are
quite similar and differences in details are not of interest for the
following discussion.

From the above assumption it is evident that the thermal speed of the
electron is higher by the ``mass'' factor
$\sqrt{m_{p}/m_{e}}\approx 43$ compared to that of the protons. The
same argument holds true for heavier ions, and thus the thermal speed
of $\alpha$-particles is half of that of the protons. The electron
impact cross-sections $\sigma^{ei}(H,He)$ are smaller by roughly a
factor 100 compared to $\sigma_{cx}(H^{+}+H\rightarrow H+H^{+})$.
Nevertheless, because the thermal speed of the electrons is, due to
the mass factor, 43 times higher than that of the protons, the rate
coefficients $\beta_{s}=\sigma_{s}(v_{rel})v_{rel}$ between those
reactions become comparable. The electron impact ionization can
dominate in the heliosheath with its high temperatures because there
the maximum of its cross-sections is about $3\cdot10^{-17}$\,cm$^{2}$
(He) or $6\cdot10^{-17}$\,cm$^{2}$ (H).   Where $v_{rel}$ is assumed
to be as follows:
\begin{equation}
v_{rel,s}^{ei} = v_{rel} = \sqrt{f_{e}\left(\frac{T_{e}}{m_{e}}+\frac{T_{s}}{m_{s}}\right)+(\vec{v}_{p=e}-\vec{v}_{H})^{2}}
\end{equation}
The index $s$ represents one of the \Hy or \He ionizing
reactions. See Appendix~B for a more thorough discussion including the
factor $f_{e}$. Note: If the relative kinetic energy
$E_{rel,s}=0.5 m_{e}(v_{rel,s}^{ei})^{2}$ is less than the ionization
energy $K_{s}$ for the species $s$ a ionization of the neutral atom
$s$ will not happen.

In the outer heliosphere the temperature plays more and more an
essential role, because the relative bulk speed between the ionized
and neutral species $|\vec{u}_{2}-\vec{u}_{1}|$ becomes small, and the
corresponding thermal speeds increase. This can easily be seen in
Fig.~\ref{Fig5}, where the ratio of the bulk speed to the thermal
speed is plotted. Moreover, only electrons with energies above the
ionization potential can ionize neutral atoms.

\section{Interaction terms\label{sec:4} }

In a multifluid multispecies (M)HD set of equations only the Euler
equations including the total density, i.e. the sum of all species
densities, and the total thermal pressure, i.e. the sum of all partial
pressures, describe the complex dynamics of the large-scale
system. Moreover, the equations for all ionized and all neutral
species need to be treated differently. This set of equations will be
called the governing equations. The governing equations include the
ram pressure and other force densities or stresses (see Appendix~A).

Each single species, ionized or neutral, needs to be followed as a
tracer fluid, e.g. their densities and thermal energy contributions to
the governing equations need to be known. This set of equations is
called the balance equations.

In most of the hydrodynamic
\citep[e.g.][]{Fahr-etal-2000, Scherer-Ferreira-2005a} or
magneto-hydrodynamic models
\citep[e.g.][]{Malama-etal-2006,Pogorelov-etal-2009b,
  Alouani-Bibi-etal-2011}
of the heliosphere the interactions between neutral \Hy and
protons can be written as balance equations and governing equations
for the dynamics \citep{Izmodenov-etal-2005}. In the balance equations
only the interaction terms are needed, which are summed over all
interactions terms and add up to zero, and the two governing
equations are the sum of all dynamically relevant ionized and neutral
species, respectively.  The ENAs are taken into account, so that the
sum over all species adds to zero. The interaction of a \PUP with a
neutral \Hy atom produces a \PUP and an ENA. Therefore, the gains and losses
in the \PUP balance equation add to zero and thus are
omitted. In the governing equations the energetic neutral atoms (ENAs)
are neglected, because it is assumed that they do not contribute to
the dynamics. 

Nevertheless, when we assume to have two distinct populations of ions
and neutrals, we must be careful about the heliospheric regions in
which the interactions happen:
\begin{itemize}
\item[1.)] Inside the termination shock, a charge exchange between H and
  p produces an \ENA$_{0}$ and a \PUI$_{0}$.  This new \PUI$_{0}$ population
  is hotter and can also interact with the neutrals, thus producing a
  new population of \ENA$_{1}$ and \PUI$_{1}$. The populations can then
  interact again with each other, which creates an hierarchy of
  populations. These second-order effects may be neglected inside the
  termination shock \citep[see e.g.\ ][]{Zank-etal-1996, Pogorelov-etal-2006}.

\item[2.)] In the heliosheath, we have a ``cooler'' population of
  shocked solar wind protons p$^{\prime}$ and a ``hotter'' shocked
  \PUI$^{\prime}$ population. Both can now interact with the neutrals and
  create an additional \ENA$^{\prime}$ populations  which now
  contributes to the original \Hy distribution and increases the
  temperature of the latter. Also the ENAs can interact with the ions
  in the heliosheath, and then produce ions which are much faster than
  the bulk speed. If picked up, they will lose energy to the plasma
  and heat it. Another plausible scenario is that this hot new
  \PUI$^{\prime}$ population, which has energies of keV, may become the seed
  population for anomalous cosmic rays. A detailed analysis of these
  heating processes will be done in a future study.

\item[3.)] Also in the outer heliosheath and beyond the bow shock
  [if it exists \citep{McComas-etal-2012, Ben-Jaffel-Ratkiewicz-2012,Zank-etal-2013,Ben-Jaffel-etal-2013,Scherer-Fichtner-2014}]
  new populations of neutrals will be generated. These new populations
  legitimate the multifluid approach for most of the MHD models
  \citep[e.g.\
  ][]{Pogorelov-etal-2009b,Alouani-Bibi-etal-2011,Izmodenov-etal-2005}.
\end{itemize}
These new populations of \PUI{}s
and \ENA{}s
need to be integrated into the governing equations (see
appendix~\ref{app:1}), while the  new balance equations are
needed to describe the individual states of the species in mind. 

To avoid a clumsy notation in the following we do not distinguish
between the different regions of the helio-/astrosphere, because the
equations are the same. Nevertheless, one should keep in mind that the
different populations, with their distinct thermodynamical states,
lead to various dynamic effects and need to be taken into account,
either by the above-mentioned multifluid approach or similar
approaches \citep{Fahr-etal-2000}.

Because we make the assumption that all involved species have the same
bulk speed as that of the corresponding governing equations, the
balance momentum equation can be neglected. Also in the energy
equation one needs only to treat the thermal energy of the species in
mind.

This assumption implies a caveat: Because of the different behavior of
the neutral hydrogen and helium fluid inside the heliopause, for
example the defocusing of hydrogen and focusing of helium close to the
Sun, these species may not be handled as one fluid in the entire
heliosphere. At least close to the Sun a kinetic treatment is
necessary \citep[e.g.\
][]{Osterbart-Fahr-1992,Izmodenov-etal-2003}. Nevertheless, for the
present study it appears as a reasonable approximation, because a
difference of a few km/s in the bulk speeds is negligible in the
relative speeds, which are determined mainly by the thermal speed in
the heliosheath. 

We discuss the individual balance terms from the general (M)HD set of
equations (see Appendix~A) with the following convention: $\sigma^{s}$
is the cross section for $s\in\{cx,ei,pi\}$ where $cx$ stands for the
charge exchange, $ei$ for electron impact, and $pi$ for
photo-ionization.
$\beta^{r}=\sigma(v_{rel,ij}^{s,\sigma})v_{rel,ij}^{s,I_{coll}}$
denotes the rate coefficient, where $r$ denotes the different relative
speeds $r=c,m,e$ for the continuity equation $c$, the momentum
equation $m$, and the energy equation, and $ij$ denote the
corresponding interaction. The superscripts $\{\sigma,I_{coll}\}$
denote the different relative speeds defined by
\citet{McNutt-etal-1998}. For details see Appendix~B. Most useful are
also the charge exchange rates $\nu_{X}^{r} = \beta^{r} n_{X}$, where
$n_{X}$ is the number density in the corresponding balance equation:
for example the balance term $S_{p}^{r}$ for protons. The balance term
for the momentum equation is a vector $\vec{S}^{r}_{X}$. The quatities
$\rho_{X}, \vec{v}_{Y}, E_{X}$ denote the density, bulk velocity, and
total energy density of species. The indices $X$ include that of the
governing equations with the indices $Y\in\{i,n\}$ for the ions $i$
and neutrals $n$ and for each species
$X\in\{$Y,p,H,H$^{+}$,H$^{0}$,...$\}$, where p are the protons, H the
neutral hydrogen atoms, H$^{+}$ the newly born ions (pick-up
hydrogen), and H$^{0}$ the newly created energetic neutral H atoms. We
denote with $w_{X}^{2}=2 \kappa T_{X} / m_{X}$ the thermal speed for
temperature $T_{X}$ with the Boltzmann constant $\kappa$ and mass
$m_{X}$.

In the following the balance equations are expressed in a form
discussed by \citet{McNutt-etal-1998} and the governing equations
as sum of the balance equations, where the ENA contribution to the
neutral component is neglected. It should also be recognized that
the equations given below are only valid for particles with the same
mass. Interaction terms between heavy and light species need a
correction term \citep[see Eq.(22) in][]{McNutt-etal-1998}. This is
discussed in more detail in Appendix~\ref{app:12}.

The balance and governing equations for the interaction between
protons, \PUP, the neutrals, and the \Hy-ENAs are explicitly stated
below. Because the indices $PUI, ENA$ are not unique and can also be
used for other ion or neutrals, we use here $\pH,\EH$ to denote the
PUI and ENAs for hydrogen.

The balance terms in the continuity equations are:
\begin{eqnarray}\label{eq:2}
S_{p}^{c}   &=& -\nu^{c}_{H}\rho_{p} -\nu^{c}_{\EH}\rho_{p} \\
S_{H}^{c}   &=& -(\nu^{pi}+\nu^{ei}+\nu^{c}_{p}+\nu^{c}_{\pH})\rho_{H}\\
S_{\pH}^{c} &=&
+(\nu^{pi}+\nu^{ei}+\nu^{c}_{p}+\nu^{c}_{\pH})\rho_{H} +\\\nonumber
&&(\nu^{pi}+\nu^{ei}+\nu^{c}_{p}+\nu^{c}_{\pH})\rho_{\EH} -
\nu_{H}^{c}\rho_{\pH} - \nu^{c}_{\EH}\rho_{\pH}\\
S_{\EH}^{c} &=& +\nu^{c}_{H}\rho_{p} -(\nu^{pi}+\nu^{ei}+\nu_{p}^{c}+\nu^{c}_{\pH})\rho_{\EH}+\nu^{c}_{\EH}\rho_{p}
\end{eqnarray}
where the dynamics is governed by the sum of protons and \PUPs for the
ions (index $i$) as well as for the neutral component (index $n$):
\begin{eqnarray}\label{eq:3}
  S_{i}^{c}   &=& S_{p}^{c}+S_{\pH}^{c}  =
  (\nu^{pi}+\nu^{ei})(\rho_{H}+\rho_{\EH}) \\\label{eq:310}
S_{n}^{c}   &=& S_{H}^{c} +s_{\EH}^{c} = 
-(\nu^{pi}+\nu^{ei}+\nu^{c}_{p}+\nu^{c}_{\pH})\rho_{H}
\end{eqnarray}  
The ENAs\EH\ are taken into account, so that the sum over all species
adds to zero. Because the ENAs produced in the solar wind have
  velocities around 400\,km/s, they will leave the system without
  further interaction. In that
  case the hydrogen population is diminished, as indicated by the last
two terms in Eq.~\ref{eq:310}. In the heliosheath where the
speeds of the  ENAs and the solar wind plasma are similar, the total
number of neutrals remains constant and the last two terms of
Eq.~\ref{eq:310} vanish. As explained above, the second order interactions with a
\PUP and a neutral
\Hy atom produces a \mbox{\PUP$^{\prime}$} and an
  \mbox{ENA\EH$^{\prime}$}, which is sorted into the \PUP and ENA\EH \
  balance equations. Therefore,
gains and losses in \PUP\ and ENA\EH\ balance equation add to zero,
e.g.:
 $\nu^{c}_{\pH}\rho_{H}=\nu^{c}_{H}\rho_{\pH}$ and
  $\nu^{c}_{\pH}\rho_{\EH}=\nu^{c}_{\EH}\rho_{\pH}$.

The balance terms in the momentum equations read:
\begin{eqnarray}\label{eq:41}
\vec{S}_{p}^{m}   &=& -(\nu^{m}_{H}+\nu^{m}_{\EH})\rho_{p}\vec{v}_{p} \\\label{eq:42}
\vec{S}_{H}^{m}   &=& -(\nu^{pi}+\nu^{ei}+\nu^{m}_{p}+ \nu^{m}_{\pH})\rho_{H} \vec{v}_{H}\\\label{eq:43}
\vec{S}_{\pH}^{m}  &=& +(\nu^{pi}+\nu^{ei}+\nu^{m}_{p}+\nu_{\pH}^{m})\rho_{H} \vec{v}_{H} \\\nonumber
           &&+(\nu^{pi}+\nu^{ei}+\nu^{m}_{p}+\nu^{m}_{\pH})\rho_{\EH}\vec{v}_{\EH} \\\nonumber
           &&-(\nu^{m}_{H}+\nu^{m}_{\EH})\rho_{\pH}\vec{v}_{\pH} \\\label{eq:44}
\vec{S}_{\EH}^{m}  &=& +
(\nu^{m}_{H}+\nu^{m}_{\EH})\rho_{p}\vec{v}_{p}\\\nonumber
           &&-(\nu^{pi}+\nu^{ei}+\nu^{m}_{p}+\nu^{m}_{\pH})\rho_{\EH}\vec{v}_{\EH}\\\nonumber
           && -(\nu^{m}_{p}+\nu^{m}_{\pH})\rho_{\EH}\vec{v}_{\EH} 
\end{eqnarray}
Here the exchange between the \PUPs and \Hy atoms must be taken
into account, because the momentum loss of the \PUP is not balanced by
the momentum gain from the \Hy. 

As discussed above, we assume that the bulk velocities of all charged
and neutral species are $\vec{v}_{i}=\vec{v}_{p}=\vec{v}_{\EH}$ and
$\vec{v}_{n}=\vec{v}_{H}=\vec{v}_{\pH}$. This assumption does not hold
everywhere, for example inside the termination shock of the
heliosphere the velocities of the newly created \ENA{}s
are high and they escape the system with out further interaction. In
the heliosheath, where the bulk velocities of the charged and neutral
particles are comparable, the \ENA\ contribution to the \Hy population
shall not be neglected. For a discussion to include second order
effects, see \citet{Zank-etal-1996}, \citet{Pogorelov-etal-2006},
\citet{Izmodenov-2007}, \citet{Alouani-Bibi-etal-2011}.

Then balance terms in the governing equations are:
\begin{eqnarray}\label{eq:5}
\vec{S}_{i}^{m}   &=& \vec{S}_{p}^{m}+\vec{S}_{\PUP}^{m} \\\nonumber
                &=&-(\nu^{m}_{H}+\nu_{\EH}^{m})(\rho_{p}\vec{v}_{i}+\rho_{\pH}\vec{v}_{n})\\\nonumber
                && +(\nu^{pi}+\nu^{ei}+\nu^{m}_{p}+\nu^{m}_{\pH})(\rho_{H}\vec{v}_{n}+\rho_{\pH}\vec{v}_{i})\\
\vec{S}_{n}^{m}   &=&
\vec{S}_{H}^{m}+ \vec{S}_{\EH}^{m} =
-(\nu^{pi}+\nu^{ei}+\nu^{m}_{p}+\nu^{m}_{\pH})(\rho_{H}\vec{v}_{n}
+\rho_{\EH}\vec{v}_{i}) \\\nonumber
&& + (\nu^{m}_{H}+\nu_{\EH}^{m})(\rho_{p}\vec{v}_{i}+\rho_{\pH}\vec{v}_{n})
\end{eqnarray}

For better reading, we omitted the indices for electrons and photons
in the rates $\nu^{ei}$ and $\nu^{pi}$ because they are unique.

The energy equations are a little more complicated. We first define
the total energy $E_{k}$ of a species $k$ as:
$E_{k} = P_{k}+0.5 \rho_{j}v_{j}^{2}$, where $P_{i}$ is the (partial)
thermal pressure of species $i$, and $\rho_{j}v_{j}^{2}$ the ram
pressure of either the sum of ions ($j=i$) or the sum of the neutrals
$j=n$. The thermal energy after the exchange is according to
\citet{McNutt-etal-1998}:
\begin{eqnarray}\label{eq:6}
S_{p}^{e}   &=& - 2 \nu_{H}^{e}\left(E_{H}\frac{\rho_{p}}{\rho_{H}} -
  E_{P} \right) - \frac{1}{2}\nu_{H}^{m}\rho_{p}
\frac{w_{p}^{2}}{w_{H}^{2}+w_{p}^{2}}(v_{H}^{2}-v_{p}^{2})\\\nonumber
&&  - 2 \nu_{\EH}^{e}\left(E_{\EH}\frac{\rho_{p}}{\rho_{\EH}} -
  E_{P} \right) - \frac{1}{2}\nu_{\EH}^{m}\rho_{p}
\frac{w_{p}^{2}}{w_{\EH}^{2}+w_{p}^{2}}(v_{\EH}^{2}-v_{p}^{2}) \\
\end{eqnarray}
\begin{eqnarray}
S_{H}^{e}   &=& -(\nu^{pi}+\nu^{ei})E_{H}  -2
\nu_{p}^{e}\left(E_{p}\frac{\rho_{H}}{\rho_{p}} -   E_{H} \right) \\\nonumber 
&&-\frac{1}{2}\nu_{p}^{m}\rho_{H}
\frac{w_{H}^{2}}{w_{p}^{2}+w_{H}^{2}}(v_{p}^{2}-v_{H}^{2})\\\nonumber
&&  -(\nu^{pi}+\nu^{ei})E_{\EH}  -2 \nu_{\pH}^{e}\left(E_{\pH}\frac{\rho_{H}}{\rho_{\pH}} -
  E_{H} \right) \\\nonumber
&&- \frac{1}{2}\nu_{\pH}^{m}\rho_{H}
\frac{w_{H}^{2}}{w_{\pH}^{2}+w_{H}^{2}}(v_{\pH}^{2}-v_{H}^{2})\\
\end{eqnarray}
\begin{eqnarray}
S_{\pH}^{e}   &=& (\nu^{pi}+\nu^{ei})E_{H}  + 2 \nu_{p}^{e}\left(E_{p}\frac{\rho_{H}}{\rho_{p}} -
  E_{H} \right) + \\\nonumber
&&\frac{1}{2}\nu_{p}^{m}\rho_{H}
\frac{w_{H}^{2}}{w_{p}^{2}+w_{H}^{2}}(v_{p}^{2}-v_{H}^{2})\\\nonumber
&&+(\nu^{pi}+\nu^{ei})E_{\EH}  + 2 \nu_{p}^{e}\left(E_{p}\frac{\rho_{\EH}}{\rho_{p}} -
  E_{\EH} \right) \\\nonumber
&&+ \frac{1}{2}\nu_{p}^{m}\rho_{\EH}
\frac{w_{\EH}^{2}}{w_{p}^{2}+w_{\EH}^{2}}(v_{p}^{2}-v_{\EH}^{2})\\\nonumber
&&- 2 \nu_{H}^{e}\left(E_{H}\frac{\rho_{\pH}}{\rho_{H}} -
  E_{\pH} \right) - \frac{1}{2}\nu_{H}^{m}\rho_{\pH}
\frac{w_{\pH}^{2}}{w_{H}^{2}+w_{\pH}^{2}}(v_{H}^{2}-v_{\pH}^{2})\\\nonumber
&&  - 2 \nu_{\EH}^{e}\left(E_{\EH}\frac{\rho_{\pH}}{\rho_{\EH}} -
  E_{\pH} \right) - \frac{1}{2}\nu_{\EH}^{m}\rho_{\pH}
\frac{w_{\pH}^{2}}{w_{\EH}^{2}+w_{\pH}^{2}}(v_{\EH}^{2}-v_{\pH}^{2}) \\\nonumber
\end{eqnarray}
\begin{eqnarray}
S_{\EH}^{e}   &=&+ 2 \nu_{H}^{e}\left(E_{H}\frac{\rho_{p}}{\rho_{H}} -
  E_{P} \right) + \frac{1}{2}\nu_{H}^{m}\rho_{p}
\frac{w_{p}^{2}}{w_{H}^{2}+w_{p}^{2}}(v_{H}^{2}-v_{p}^{2})\\\nonumber
&&  + 2 \nu_{\EH}^{e}\left(E_{\EH}\frac{\rho_{p}}{\rho_{\EH}} -
  E_{P} \right) + \frac{1}{2}\nu_{\EH}^{m}\rho_{p}
\frac{w_{p}^{2}}{w_{\EH}^{2}+w_{p}^{2}}(v_{\EH}^{2}-v_{p}^{2}) \\\nonumber
&&-(\nu^{pi}+\nu^{ei})E_{\EH}  - 2 \nu_{p}^{e}\left(E_{p}\frac{\rho_{\EH}}{\rho_{p}} -
  E_{\EH} \right) \\\nonumber
&&- \frac{1}{2}\nu_{p}^{m}\rho_{\EH}
\frac{w_{\EH}^{2}}{w_{p}^{2}+w_{\EH}^{2}}(v_{\pH}^{2}-v_{\EH}^{2})\\\nonumber
&&+ 2 \nu_{H}^{e}\left(E_{H}\frac{\rho_{\pH}}{\rho_{H}} -
  E_{\pH} \right) + \frac{1}{2}\nu_{H}^{m}\rho_{\pH}
\frac{w_{\pH}^{2}}{w_{H}^{2}+w_{\pH}^{2}}(v_{H}^{2}-v_{\pH}^{2})\\\nonumber
&&  + 2 \nu_{\EH}^{e}\left(E_{\EH}\frac{\rho_{\pH}}{\rho_{\EH}} -
  E_{\pH} \right) + \frac{1}{2}\nu_{\EH}^{m}\rho_{\pH}
\frac{w_{\pH}^{2}}{w_{\EH}^{2}+w_{\pH}^{2}}(v_{\EH}^{2}-v_{\pH}^{2}) \\\nonumber
\end{eqnarray}
where we have indicated with the superscript $m,e$ that the relative
speeds from the collision integrals for the momentum and energy
equation have to be taken (see Appendix~B). Moreover, with the above
assumption $\vec{v}_{p}=\vec{v}_{\EH}$ and $\vec{v}_{H}=\vec{v_{\pH}}$
some terms in the above energy balance equation vanish.

For the governing energy equations we have to take care of the change
in the total energy, which adds the following terms: 
\begin{eqnarray}
  S_{i}^{tot} &=& \frac{1}{2}\nu_{p}^{m}(v_{H}^{2}-v_{p}^{2})\rho_{H}
  \left(\frac{w_{H}^{2}-2w_{p}^{2}}{w_{H}^{2}+w_{p}^{2}} \right)\\\nonumber
&& + 
\frac{1}{2}\nu_{\pH}^{m}(v_{\EH}^{2}-v_{\pH}^{2})\rho_{\EH}
  \left(\frac{w_{\EH}^{2}-2w_{\pH}^{2}}{w_{\EH}^{2}+w_{\pH}^{2}}
  \right)\\
  S_{n}^{tot} &=& \frac{1}{2}\nu_{H}^{m}(v_{p}^{2}-v_{H}^{2})\rho_{p}
  \left(\frac{w_{p}^{2}-2w_{H}^{2}}{w_{p}^{2}+w_{H}^{2}} \right) \\\nonumber
&&+ 
\frac{1}{2}\nu_{\EH}^{m}(v_{\pH}^{2}-v_{\EH}^{2})\rho_{\pH}
  \left(\frac{w_{\pH}^{2}-2w_{\EH}^{2}}{w_{\pH}^{2}+w_{\EH}^{2}}
  \right)
\end{eqnarray}
because of $\vec{v}_{p}=\vec{v}_{\EH}$ and $\vec{v}_{H}=\vec{v_{\pH}}$
all other contributions vanish.

Then the balance terms in the governing energy equations are:

\begin{eqnarray}\label{eq:7}
S_{i}^{e}   &=& S_{p}^{e}+S_{\pH}^{e}+S_{i}^{tot}\\
S_{n}^{e}   &=& S_{H}^{e}+S_{\EH}^{e}+S_{n}^{tot}
\end{eqnarray}
To save space we waived the explicit sums.

The photo-ionization rate is given by
\begin{equation}\nonumber
\nu^{pi} = \left<\sigma^{pi} F_{EUV}\right>_{0} \dfrac{r_{0}^{2}}{r^{2}}=
8\cdot10^{-8}\dfrac{r_{0}^{2}}{r^{2}} [s^{-1}],
\end{equation}
{with $r_{0}=1$\,AU. For the electron impact rates,
$\nu^{ei} = n_{e} \sigma(E_{e}) v_{rel,e}^{m,n}= n_{e} \beta_{e}$
is assumed and for the electron number density
$n_{e} = (n_{p}+n_{\PUP})$ (quasi neutrality).  For the relative speed
$v_{rel,e}^{m,n}$ a similar approach is used as in appendix~B. Because
of the high thermal speeds of the electrons (assuming thermal
equilibrium between ions and electrons) the relative speed is roughly
equal to the thermal electron speed. As discussed in
\citet{Fahr-Chalov-2013} the electron temperature may increase further
downstream the termination shock.

Usually all interaction terms with the newly created PUI{}s
and ENA{s} are neglected, i.e. all terms in the above equations
containing terms with H+H$^{+}$, p+H$^{0}$ and H$^{+}$+H$^{0}$ are set
to zero. These assumptions may not hold beyond the termination shock
where the speeds of charged and neutral particles can be in the same
order depending on the location.

Other forms of the interaction terms are discussed by
\citet{Williams-etal-1997}. A detailed description of the relative
speeds is given in appendix~B.

To include a new species into the model, the above equations have
to be extended: for each species a separate set of Euler equations
is required. Differences in the ion velocities are assumed to be
equalized on a kinetic scale by wave-particle interactions. Thus, for
the much larger MHD scales we will assume that all ion
velocities are equal to the bulk velocity of the main fluid.  

Therefore, the dynamics will be governed by the above equations with
the indices $i,n$, where now the additional species need to be
added. Including \He, the set of the balance equations consists of
those for H, He, H$^{+}$, He$^{+}$, He$^{2+}$, and the governing
equations for ions and neutrals. In the latter the total density
$\rho_{i}$ is the sum of all ionized species, i.e.\
$\rho_{i}=\rho_{H^{+}}+\rho_{He^{+}}+\rho_{He^{2+}}$ and
$\rho_{n}=\rho_{H}+\rho_{He}$ for the neutral species. The governing
momentum equation for $i,n$ describes the bulk velocity of the system,
where the pressure term is the sum of the partial pressures of all
relevant species.  The energy equations have to be treated
analogously. The momentum equations for the individual species can be
neglected far away from obstacles, because we assume that due to
wave-particle interactions all  bulk velocity differences shall
vanish. Close to obstacles, like the Sun/star or a shock front, the
individual bulk velocities of the species may differ by the Alfv\'en
speed, as observed in the solar wind for $\alpha$-particles
\citep{Marsch-etal-1982,Gershman-etal-2012} and L. Berger (private
communication). Because the Alfv\'enic disturbances travel along the
magnetic field, which is assumed to be the Parker-spiral, the radial
speed in the outer heliosphere is the bulk speed for all
species. Nevertheless, these particles can induce a perpendicular
pressure as well as diamagnetic effects as described for \PUPs
\citep[see][]{Fahr-Scherer-2004b,Fahr-Scherer-2004c}. A thorough
discussion of the effects of $\alpha$-particles in the heliosheath
goes far beyond the scope of this paper, and will be studied in future
work.

For all practical purposes one needs to solve the governing equations,
including the continuity and energy balance equations of all other
species, because they can influence the dynamics in the inner
heliosheath and, more general, that of astrosheaths.

The above set of equations concerning the interaction terms have now
to be extended to include the charge exchange between ionized and
neutral \Hy and/or \He atoms, as well as the electron impact
reactions for both neutral species. A complete set of
equations can be found in appendix~\ref{app:2} in the
tables~\ref{tab:1} and~\ref{tbb}.

Since, the solar abundance of $\alpha$-particles is 4\% \citep{Lie-Svendsen-etal-2003}
of that of the protons, and the singly-charged \He is even less
abundant, we neglect in the following the interaction with ionized
solar wind \He and the neutral interstellar \Hy or \He
atoms. For other astrospheres these abundances can change and also the
stellar wind speed can be much higher, in that case the charge
exchange and electron impact cross-sections discussed below can have
completely different relevance.

All the reactions discussed in Appendix~C can play an important role
depending on the astro-/heliosphere model. 

For example, astrospheres can have bulk velocities in the range of a
few thousand km/s and hence their relative energies per nucleon
can be in the ten keV range, where reactions, like
\mbox{He$^{2+}$+H$\rightarrow$H$^{+}$+He$^{+}$} become important
see Fig.~\ref{Fig2} or tables~\ref{tab:1} and~\ref{tbb}, which can
be neglected in the heliosphere. In the following section we will
concentrate on the electron impact ionization of H and He and show
that they play an non-negligible role in the inner heliosheath and
inside the termination shock.

Note that including species others than \Hy in charge exchange
processes can change the number density of electrons, for example 
He$^{+}$+He$^{+}$$\rightarrow$He$^{+}$+He$^{2+}$+e. This must be taken
care off in equations containing the electron number density (not
discussed here).

\section{Electron impact ionization of H and He\label{sec:5} }

\begin{figure*}
  \includegraphics[width=1.\columnwidth]{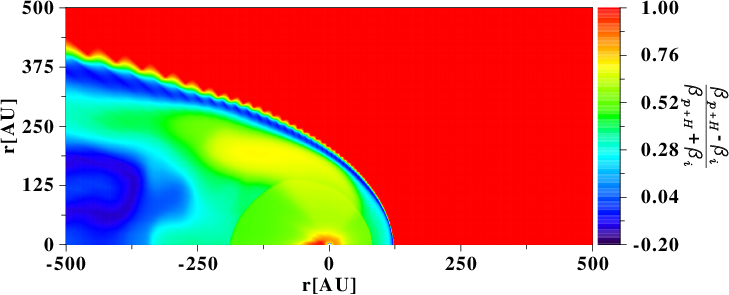}\vertihori
  \includegraphics[width=1.\columnwidth]{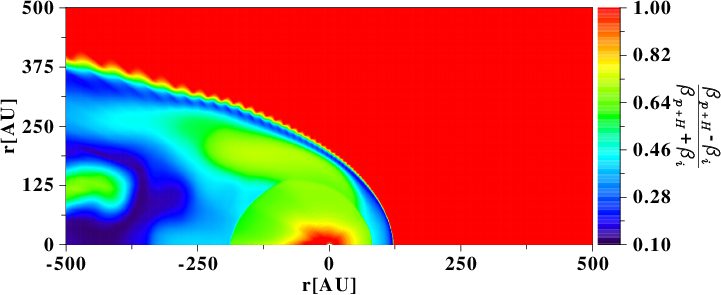}
  \caption{The ratios as defined in Eq.~\ref{eq:14}. The left panel shows
    that for \Hy while the right panel presents the ratio for
    singly ionized \He. Positive values indicate, that the charge
    exchange process (H+H$^{+}\rightarrow$H$^{+}$+H) dominates, while
    for negative values the electron impact is more relevant. Note the
    different scales in the color bars.
  \label{Fig4}}%
\end{figure*}
The cross-section for the electron impact reactions were already shown
in Fig.~\ref{Fig2}. To demonstrate their importance we will estimate
their contribution in the governing continuity equations. They can be
written in the following form (see Eq.~\ref{eq:3}):
\begin{align}\label{eq:9a}
S_{i}^{c}   &= S_{p}^{c}+S_{\pH}^{c}+S_{He^{+}}^{c} + S_{He^{++}}^{c}
= + (\nu^{pi}+\nu^{ei})\rho_{H}  \\\nonumber
&+(\nu^{pi}_{He^{+}} +\nu^{ei}_{He^{+}}+\nu^{pi}_{He^{++}}
+\nu^{ei}_{He^{++}}) \rho_{He}\\
\label{eq:9b}
 S_{n}^{c}   &= S_{H}^{c}+S_{He}^{c}= -(\nu^{pi}+\nu^{ei}+\nu^{c}_{p})\rho_{H} \\\nonumber
&- (\nu^{pi}_{He^{+}} +\nu^{ei}_{He^{+}}+\nu^{pi}_{He^{++}} +\nu^{ei}_{He^{++}}) \rho_{He}
\end{align}
where we have neglected all charge exchange reactions between H and
He, both neutral or ionized, because the rate coefficients $\beta_{i}$
are less then a factor 10$^{-3}$ of those for to electron impact or
photo-ionization. The only comparable rate-coefficient is that between
protons and neutral \Hy atoms. For the corresponding balance equation
see appendix~A. The notation $\nu^{ei}_{He^{+}}$ or $\nu^{ei}_{He^{++}}$
indicates the ionization of neutral \He into singly and doubly ionized
\He.  Moreover, we neglected all the higher order interaction between
\PUI{}s and \ENA{}s. 

Also care must be taken, when the interstellar ionized \He is
  taken into account. The abundances of singly-charged \He can be of
  the same order as those of the neutral component
  \citep{Wolff-etal-1999}.  The incidence of He$^{++}$ in the
  interstellar medium are modeled \citep{Slavin-Frisch-2008} and are
  negligible according to the model.

First we assume that the interstellar abundances of \He is 10\%
that of \Hy \citep{Asplund-etal-2009}, but see also
\citet{Moebius-etal-2004} and \citet{Witte-2004} for the determination
of the \He content based on observations inside 5\,AU. Furthermore,
we approximate the mass of \He $m_{He}$ to be four times that of
the \Hy mass $m_{H}$. Then we can write for the region inside the heliopause:
\begin{equation}
  \label{eq:10}
  \rho_{He} = m_{He} n_{He} = 4 m_{H} 0.1 n_{H} = 0.4 \rho_{H}
\end{equation}
The consequences for the governing equations are discussed in Appendix~A.

Furthermore, the photo-ionization rate at 1\,AU for \Hy and
\He at solar minimum is roughly $8\cdot10^{-8}$s$^{-1}$ \citep
{Bzowski-etal-2012,Bzowski-etal-2013}.  With these assumptions, we can rewrite
Eq.~\ref{eq:9a} and~\ref{eq:9b}:
\begin{align}
  \label{eq:11a}
  S_{i}^{c}  & = + \left(\nu^{pi}+\nu^{ei} + 0.4 \left[\nu^{ei}_{He^{+}}+\nu^{ei}_{He^{++}}\right]\right) \rho_{H}\\
S_{n}^{c}   &=-\left(\nu^{pi}+\nu^{ei}+\nu^{c}_{p} + 0.4 \left[\nu^{ei}_{He^{+}}+\nu^{ei}_{He^{++}} \right]\right) \rho_{H}  \label{eq:11b}
\end{align}
To determine the above rates, we need to estimate the relative
velocities. Because, to our knowledge, the electron and \He
temperatures in the heliosheath are neither observed nor modeled, we
assume that the electron temperature $T_{e}$ and \He temperature
are the same as the proton temperature $T_{p}$, i.e.\
$T_{p}=T_{e}=T_{He^{+}}=T_{He^{2+}}=10^{6}$\,K. The temperatures in
the interstellar medium for the neutral \Hy $T_{H}$ and \He
$T_{He}$ are of the order of 8000\,K, but the exact values do not play
a role, because only the sum of the ion and neutral temperature
determines the relative speeds for protons $v_{rel,p}$ and
electrons $v_{rel,e}$. For our estimation we assumed that the relative
speeds $v_{rel,Hp}^{r,\sigma}=v_{rel,Hp}^{r,I_{coll}}=v_{rel,p}$. Furthermore, we suppose that in the heliosheath
the relative bulk speeds are
$\sqrt{(\vec{v}_{p,e}-\vec{v}_{H,He})^{2}} \approx 50$\,km/s, and
neglecting the temperature of the neutrals (see above), we get:
\begin{eqnarray}\nonumber
  v_{rel,p} &=& \sqrt{ \frac{128 k_{B}}{9 \pi m_{p}} T_{p} + 50^{2}} =
    \sqrt{192^{2}+50^{2}} \approx 200\, \mathrm{km/s}
    \\\nonumber
v_{rel,e} &=& \sqrt{\frac{m_{p}}{m_{e}}} v_{rel,p}\approx 8250\,\mathrm{km/s} \approx 10^{4}\, \mathrm{km/s}
\end{eqnarray}

The relative speed of the electrons corresponds to an energy $E=0.5
m_{e}v_{rel,e}^{2}\approx 200$\,eV and we can then read from Fig.~\ref{Fig3} that the
electron impact cross-section to produce singly or doubly ionized \He
differs roughly by a factor 4, and that of singly ionized \He is
roughly the same as for \Hy. Thus
$\sigma^{ei}(H^{+}) \approx 2\sigma^{ei}_{He^{+}} \approx
6\sigma_{He^{++}}\approx 6\cdot10^{-17}$\,cm$^{2}$.
From Fig.~\ref{Fig2} we get the charge exchange cross-section
$\sigma^{cx}(H^{+}+H)\approx 10^{-15}$\,cm$^{2}$. The electron density
in the inner heliosheath is assumed to be
$\rho_{p+\pH}=\rho_{e}= 3\cdot 10^{-3}$\,cm$^{-3}$.  With these
estimates we get for the different rates in that region:
\begin{alignat}{4}\nonumber
  \nu^{pi} (100 AU) &=&8\cdot10^{-8}\frac{r_{0}^{2}}{r^{2}}&&&\approx& 0.8\cdot 10^{-11} [\mathrm{s}^{-1}]\\\nonumber
  \nu^{c}_{H}  &=& \rho_{p} \sigma^{cx} v_{rel,p} &&&\approx&
  6\cdot10^{-11}[\mathrm{s}^{-1}]\\\nonumber
\nu^{ei}_{H^{+}} &=& \rho_{e} \sigma^{ei} v_{rel,e} &&&\approx& 
  6\cdot10^{-11}[\mathrm{s}^{-1}]\\\nonumber
\nu^{ei}_{He^{+}} &=& \rho_{e} \sigma^{ei} v_{rel,e} &\approx&\frac{1}{2} \nu^{ei}_{H^{+}}&\approx& 
  3\cdot10^{-11}[\mathrm{s}^{-1}]\\\nonumber
\nu^{ei}_{He^{++}} &=& \rho_{e} \sigma^{ei} v_{rel,e} &\approx&\frac{1}{12}\cdot \nu^{ei}_{H^{+}}&\approx& 
  0.5\cdot10^{-11}[\mathrm{s}^{-1}]\\\nonumber
\end{alignat}
see \mbox{www.cbk.waw.pl/$\sim$jsokol/solarEUV.html} for the
  photo-ionization rates of H, He, Ne, O and He$^{+}$.
From the above it is evident that all the rates are of the same
order. Now Eqs~\ref{eq:11a} and~\ref{eq:11b} can be written as:
\begin{eqnarray}
  \label{eq:12a}
    S_{i}^{c}  & =& + \left(\nu^{pi}+\nu^{ei}_{H^{+}} + 0.2
      \nu^{ei}_{H^{+}}\right) \rho_{H} \\\nonumber 
&=& \nu^{pi}\rho_{H}+1.2\nu^{ei}_{H^{+}}\rho_{H}\\\label{eq:12b}
S_{n}^{c}   &=&-\left((\nu^{pi}+\nu^{ei}+\nu^{c}_{p} + 0.2
  \nu^{ei}_{H^{+}} \right) \rho_{H}\\\nonumber  &=& \nu^{pi} \rho_{H} +
\nu^{c}_{p}\rho_{H} + 1.2 \nu^{ei}\rho_{H}
\end{eqnarray}
From Eq.~\ref{eq:12a} we learn that the mass loading in the inner
heliosheath for the combined He and H electron impact interaction
terms is an important effect in the inner heliosheath, and not only
that of \Hy needs to be taken into account, but \He
contributes about 20\% to the mass loading. From Eq.~\ref{eq:12b} it
is evident that within our estimation the mass loss of neutrals by
electron impact is 20\% that of the charge exchange between solar wind
protons and neutral interstellar \Hy.

For the momentum equations we get after a similar consideration:
\begin{align}\label{eq:13a}
\vec{S}_{i}^{m}   &=  -\nu^{m}_{p}\rho_{H}\vec{v}_{p} +(\nu^{pi}+ 1.2\nu^{ei}+\nu^{m}_{p})\rho_{H} \vec{v}_{H}\\
\vec{S}_{n}^{m}   & =-(\nu^{pi}+ 1.2 \nu^{ei}+\nu^{m}_{p})\rho_{H} \vec{v}_{H}
\end{align}
where we have assumed that the interstellar bulk velocity of \He is the same as
that of \Hy.

The energy balance terms become: 
\begin{align}\label{eq:14a}
S_{i}^{e}   &= 
-\nu^{e}_{p}\dfrac{\rho_{H}}{\rho_{p}}E_{p} 
+(\nu^{pi}+1.2 \nu^{ei}+\nu^{e}_{p})\dfrac{\rho_{H}}{\rho_{p}}
E_{H}\\
S_{n}^{e}   &= -(\nu^{pi}+1.2\nu^{ei}+\nu^{e}_{p})\dfrac{\rho_{H}}{\rho_{p}} E_{H}
\end{align}

In Fig.~\ref{Fig4} we present the ratios of the rates for electron
impact ionization of \Hy and \He (singly-charged). To enhance
the visibility, we normalized to the charge exchange rate
$\sigma(H^{+}+H)$ in the following form:
\begin{eqnarray}
  \label{eq:14}
  r = \frac{\nu^{c}_{p} - \nu^{ei}_{X}}{\nu^{c}_{p} +
    \nu^{ei}_{X}}
\end{eqnarray}
where $X\in\{\mathrm{H,He}\}$. The ratio $r$ is estimated from an
existing dynamic model with high speed streams over the poles. The
cross-sections are calculated from the modeled plasma parameters,
hence they are not self-consistently taken into account. Nevertheless,
it demonstrates the relative importance of electron impact
ionization in the heliosphere, which finally has to be modeled
self-consistently.  Fig.~\ref{Fig4} shows a contribution around 20\%,
as discussed above, especially in the tail region. The importance can
be seen, when estimating the Alfv\'en speed, which is inversely
proportional to the square root of the total density of charged
particles. An enhancement of 20\% in the charged density
\citep[according to the model by][]{Slavin-Frisch-2008} will lead to
approximately 10\% reduction in the Alfv\'en speed. This is important
in the recent discussion about the bow shock
\citep{McComas-etal-2012}.  It can even be seen in Fig.~\ref{Fig4}
that also inside the termination shock, electron impact ionization is
not negligible.

As can be inferred from Fig.~\ref{Fig4}, the electron impact is
significant almost everywhere inside the heliopause and dominates
close to the heliopause and in the tail region, at least for
ionization of \Hy.  The structures in the heliotail visible in
Fig.~\ref{Fig4} are caused by the previous solar cycle activities,
which are still propagating down the heliotail
\citep[see][]{Scherer-Fahr-2003a,Scherer-Fahr-2003b,Zank-Mueller-2003}. For
the above calculations, we have estimated the cross-section from the
relative velocity and temperature taken from the model, with the
assumption of thermal equilibrium. It shows also that our rough
approximations above are quite good.

The discussion shows that electron impact effects for \Hy
and \He needs to be taken into account to improve heliospheric
models, which otherwise can lead to results that differ in
extreme by 20\% (see Eqs~\ref{eq:12a} to~\ref{eq:14}).

\section{Helium loss in the inner heliosheath\label{sec:6} }

To interpret IBEX observation, for \He and other species
\citep{Bzowski-etal-2012}, it is important to know how much \He is
lost in the inner heliosheath. 
To get an idea of the order of these losses, we use the balance continuity
equation for \He:
\begin{equation}\label{eq:20}
  \frac{\partial \rho_{He}}{\partial t} + \div(\rho_{He}\vec{V}) =
  S_{He}^{c}= - (\nu^{pi} + \nu^{ei}_{He^{+}} + \nu^{ei}_{He^{++}})\rho_{He} 
\end{equation}
where we have taken from Table~\ref{tab:1} in appendix~C only the
photo-ionization to singly charged \He, and the electron impact to
both ionization states as losses. Other losses are small, even the
photo-ionization to the doubly-charged state
\citep{Rucinski-etal-1996}. In the heliosheath the plasma is subsonic,
and with the usual assumption of incompressibility for subsonic flows
the divergence term may be neglected. Nevertheless, because due to the
ionization some particles are lost, this assumption holds rigorously
no longer true. If we assume for the moment, that we can neglect the
divergence, then Eq.~\ref{eq:20} has the solution
\begin{equation}
  \label{eq:21}
  \rho_{He} = \rho_{0} e^{-t/\tau}
\end{equation}
with $\tau^{-1} = \nu^{pi} + \nu^{ei}_{He^{+}} +
\nu^{ei}_{He^{++}}$.
For the following estimation, we further assume a lower limit for the
photo-ionization rate at 1\,AU $\nu^{pi}(1AU)=8\cdot10^{-8}$\,s$^{-1}$
\citep{Rucinski-etal-2003} and that it decreases like $r^{-2}$, the
heliosheath has a width of 40\,AU, and that the photo-ionization rate
inside the heliosheath has a constant value
$\nu^{pi}(100AU)=8\cdot10^{-12}$\,s$^{-1}$. Since the total temperature in
the heliosheath is $10^{6}$\,K (see \citet{Livadiotis-etal-2011},
but also \citet{Richardson-etal-2008} for a lower proton temperature)
then with the values for the electron impact as discussed above, we
get $\tau \approx 300$\,years. Particles with speeds of 20,25,30\,km/s
need $\approx$ 9.45, 7.56, and 6.29 years to travel through the inner
heliosheath. Inserting these numbers into Eq.~\ref{eq:21} leads to a
decrease of 2-3\% over that distance. Increasing the distance
between the heliopause and termination shock, i.e.\ for higher
latitudes or different solar activity, these numbers will increase
approximately linearly.

This loss is of the same order as that determined  by \citet{Cummings-etal-2002},
who discussed, in view of anomalous cosmic  ray composition, a loss of
\He in the heliosheath in the order of 5\% in the inner
heliosheath. Obviously for more extended astrosheaths these losses may
be even more pronounced.

\section{Charge exchange with particles of different masses\label{sec:cx}}

For the interaction of particles with different mass, the approach
by \citet{McNutt-etal-1998} can be applied. The  calculation of
the collision integrals between two species requires the knowledge of
the functional form of the charge-exchange cross-section, i.e.\ one
has to solve integrals of the form:
\begin{eqnarray}
  \label{eq:i1}
  I_{\mathrm{coll}}^{c,e} \propto
  \int\limits^{\infty}_{0}f_{1}f_{2}g^{k}\sigma^{cx}(g)dg\\\label{eq:i12}
 I_{\mathrm{coll}}^{c,e} \propto
  \int\limits^{\infty}_{0}f_{1}f_{2}g\vec{g}\sigma^{cx}(g)dg
\end{eqnarray}
where $|\vec{g}|= |\vec{V}_{2}-\vec{V}_{1}|$ is the modulus of the  velocities
$\vec{V}_{1},\vec{V}_{2}$ of the individual particles of species~1
or~2, and the indices $c,m,e$
indicate the integrals for the continuity, momentum and energy
equation (see Eq.~\ref{ap:1}), $k\in\{1,3\}$ for $c,e$, respectively. The collision integrals are equivalent
to the balance terms $S^{c,e}$ and , which are their solution under
simplifying assumptions, for an example see below.

\begin{figure}
  \includegraphics[width=0.95\columnwidth]{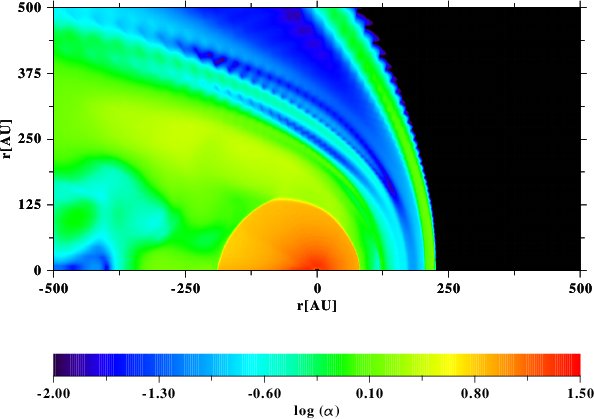}
  \caption{The $\alpha$ parameter throughout the
    heliosphere. $\alpha$ is the ratio of the modulus of the relative
    bulk velocity to the thermal speed of both species. see Eq.~\ref{eq:111}. \label{Fig5}}
\end{figure}
In most derivations it is assumed that $\sigma^{cx}(g)$ is nearly
constant
\citep{Heerikhuisen-etal-2008,McNutt-etal-1998,Alouani-Bibi-etal-2011}. This
is in general not the case, even for $\sigma^{cx}(H+p)$ this holds
only below energies of 5\,keV. Unfortunately, the integrals
$ I_{\mathrm{coll}}^{c,m,e}$ can only be solved analytically for
polynomial functional dependence of $\sigma^{cx}$. The way out of this
dilemma was sketched by \citet{McNutt-etal-1998} by developing
$\sigma^{cx}$ into a Taylor series and assuming that after a given
order all higher order terms vanish. \citet{McNutt-etal-1998} required
that only the zeroth order is relevant, and determined the
characteristic speed ($g_{0}^{c,m,e}\equiv u^{cx}_{c,m,e}$) at which
the cross-sections should be taken. The authors calculated the
characteristic speed by putting the first order of the Taylor expansion
of $\sigma^{cx}$ into the integrals $I_{\mathrm{coll}}^{c,m,e}$ and
requiring that the sum of these integrals vanishes. The latter
allows to determine a characteristic speed, following mathematically
from the mean value theorem for integrals. Nevertheless, the above
procedure holds only as long as the cross-sections are weakly varying. In
general, this is not true and as correction the higher order terms
shall be used (see appendix~\ref{app:12}).

Moreover, the relative velocities calculated for the collision
integrals $I_{\mathrm{coll}}^{c,m,e}$ are mutually distinct and also
not identical with the characteristic speeds discussed above, see also
appendix~\ref{app:12}.  With a slightly different notation introduced
in appendix~\ref{app:12} compared to the above cited paper, the
dimensionless parameter $\alpha$ reads:
\begin{eqnarray}
  \label{eq:111}
\kappa = \frac{1}{w_{1}^{2}+w_{2}^{2}}; \qquad
 \alpha =\sqrt{(\vec{v}_{2} - \vec{v}_{1})^{2}}\sqrt{\kappa}
\end{eqnarray}
with the individual particle velocities $\vec{v}_{1},\vec{v}_{2}$ and
thermal speeds $w_{1},w_{2}$ for the particles of species~1 or~2, respectively.
Then one finds that the above mentioned different speeds normalized to
the thermal speeds are functions of $\alpha$ only, i.e.
$\tilde{u}_{i}^{j}(\alpha) = \sqrt{\kappa}u_{i}^{j}$, where
$j\in\{c,m,e,P\}$ for the continuity, momentum and energy equations
and $j=P$ for the thermal pressure, respectively. The index
$i\in\{cx,rel\}$ are the speeds needed for the characteristic speeds
in the charge exchange cross-section and the corresponding relative
speeds, respectively. All speeds $u_{i}^{j}$ are called ``collision''
speeds in the following. The explicit formulas are stated in
appendix~\ref{app:12}. 

In Fig.~\ref{Fig5} the parameter $\alpha$ is presented throughout the
heliosphere. It can be seen that the $\alpha$ parameter ranges from
values near zero close to the heliopause, in the tail region and in
the outer heliosheath, where the relative velocities are small, but
the thermal ones high, to values in the range of 30 inside the
termination shock, where the relative velocity is high and the thermal
speed low. In the left panel of Fig.~\ref{Fig6} the dependence of the
$u_{j}^{i}$ from $\alpha$ is shown and in the right panel of
Fig.~\ref{Fig6} the relative ``error''
$f_{i}= (u_{i}^{rel}-u_{i}^{cx})/u_{i}^{rel}$ is presented.

It can be seen that the collision speeds for the continuity equations
nicely follow the approximation
$u_{c}^{rel}\approx u_{c}^{cx}$ for $\alpha>1$, while the speeds
$u_{m,e}^{cx}$ and $u_{m,e}^{rel}$ for the momentum and energy
equation, respectively, differ strongly for all values of $\alpha$, as
well as those for the continuity equations for small $\alpha$, as can
be nicely seen in the right panel of Fig.~\ref{Fig6}.

\begin{figure*}
  \includegraphics[width=0.95\columnwidth]{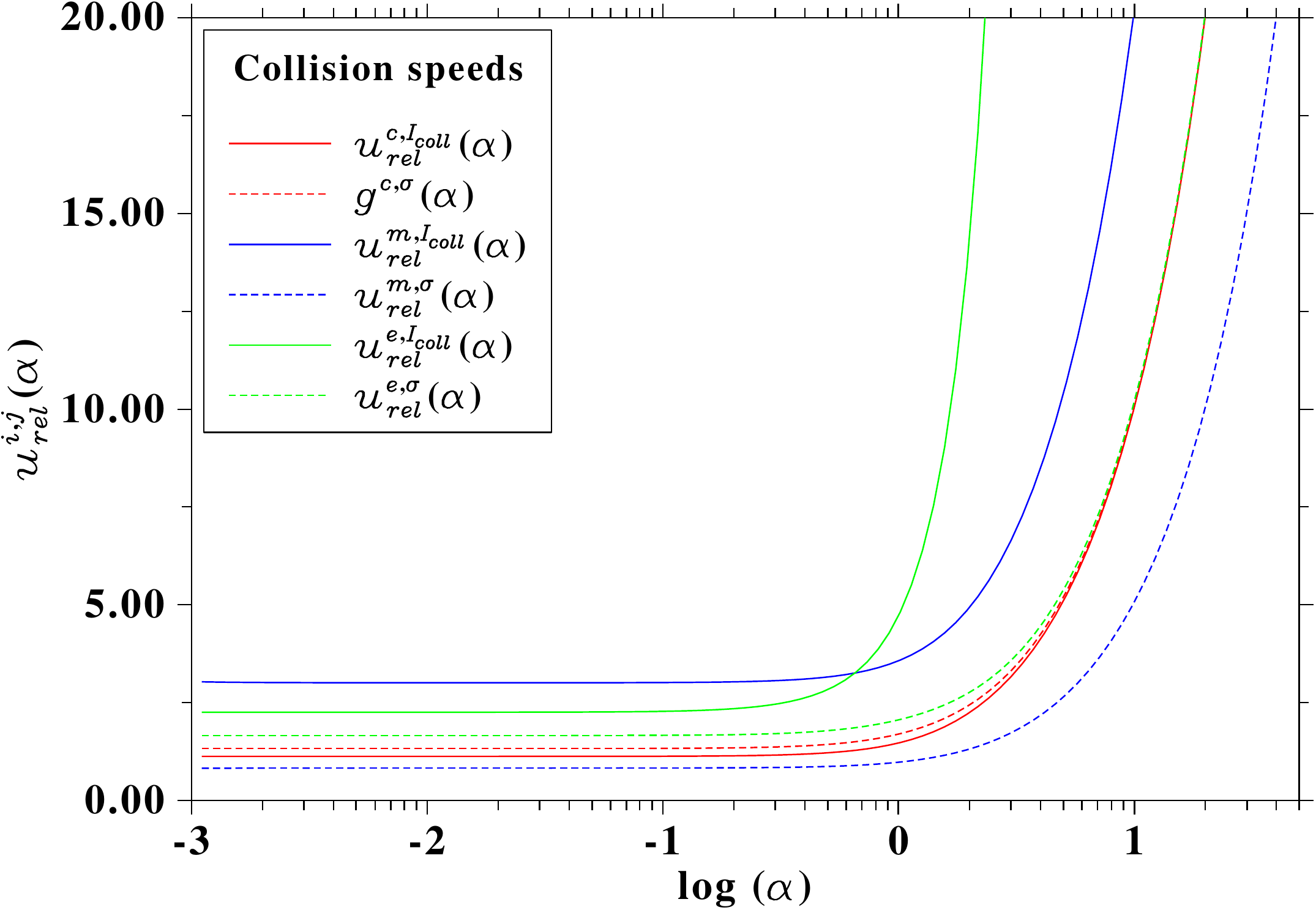}
  \includegraphics[width=0.95\columnwidth]{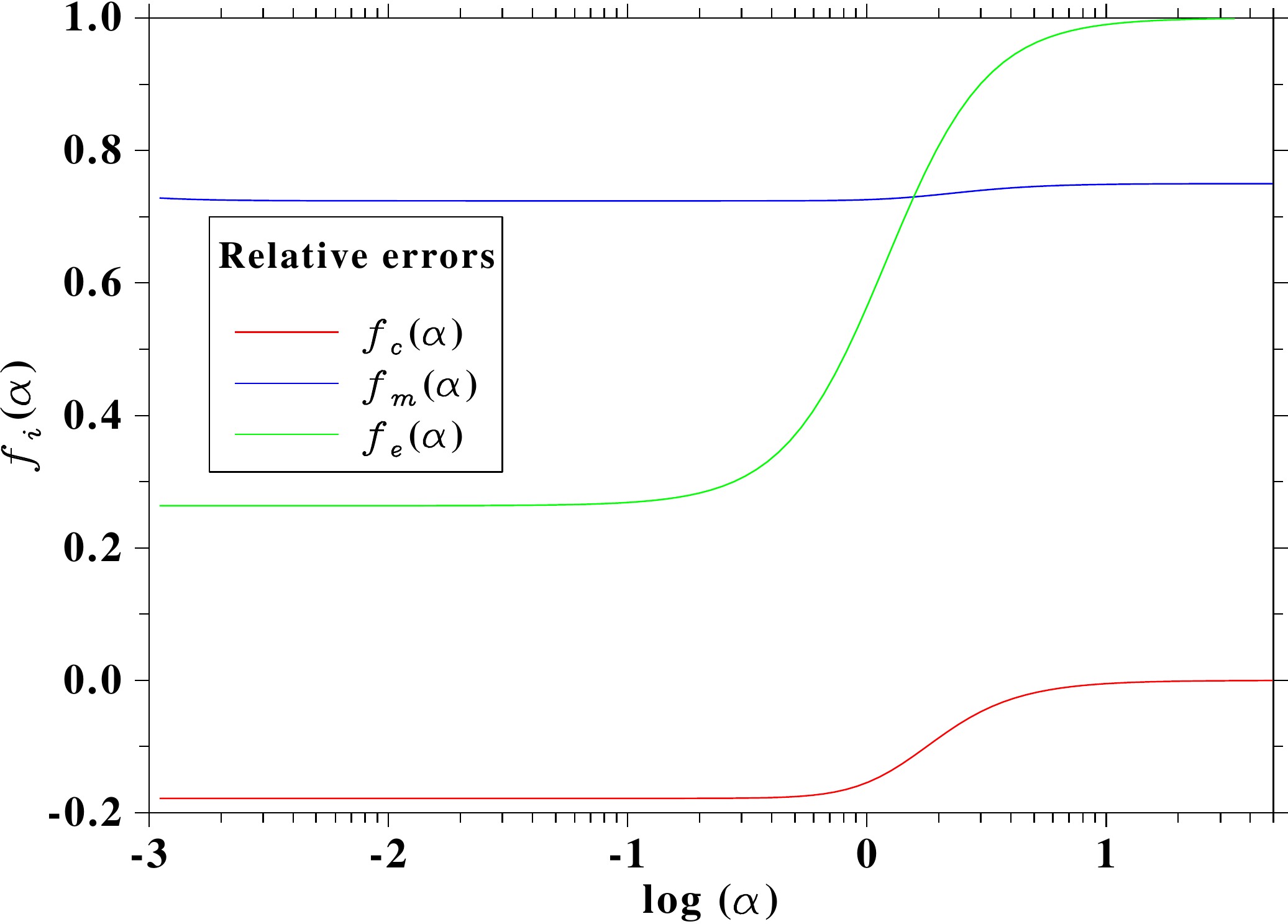}
  \caption{The characteristic collision speeds normalized to the
    thermal speed as function of the parameter $\alpha$ (left panel)
    and the relative ``error''
    $f_{r}=
    (u_{rel}^{r,I_{coll}}-u_{rel}^{r,\sigma})/u_{rel}^{r,I_{coll}}$, $r\in\{c,m,e\}$.  }\label{Fig6}
\end{figure*}

Given the above speeds the charge exchange terms read:
\begin{align}
  S_{i}^{c} &= \pm \frac{\rho_{1}\rho_{2}}{m_{i}}   \sigma^{cx}(u_{c}^{cx}) u_{c}^{rel}\\
S_{i}^{m} &= \pm \frac{\rho_{1}\rho_{2}}{m_{1}+m_{2}}\sigma^{cx}(u_{m}^{cx}) u_{m}^{rel}\Delta
\vec{v}\\
S_{i}^{e}&=  \pm
\frac{\rho_{1}\rho_{2}}{m_{1}+m_{2}}
\left(\frac{2k_{B}(T_{2}-T_{1})}{m_{1}+m_{2}}\sigma^{cx}(u_{e}^{cx})u_{e}^{rel}\right.\\\nonumber
&\left. + \sigma^{cx}(u_{m}^{cx})u_{m}^{rel}(v_{1}^{2}+v_{2}^{2}) +
\frac{1}{2}\frac{w_{2}^{2}-w_{1}^{2}}{w_{1}^{2}+m_{2}^{2}}\sigma^{cx}(u_{m}^{cx})u_{m}^{rel}(\Delta v)^{2}\right)
\end{align}

In the above formulas $i \in \{1,2\}$ stands for the respective
particle species, and the $\pm$ sign has to be chosen in such a way
that a loss in one species (for example ions) is a gain in another
one (for example fast neutrals).

Even when the neutrals are treated kinetically
\citep{Izmodenov-etal-2005,Heerikhuisen-etal-2008}, the moments of the
collision integrals must be calculated, because the ions are handled
with an MHD approach, in which implicitly the solution of those for
Maxwellian-distributed particles are modeled. Thus, the above
calculations have to be repeated with a different velocity
distribution, like a $\kappa$-distribution
\citep{Heerikhuisen-etal-2008} to obtain the required collision
speeds.

Moreover, as discussed briefly in appendix~\ref{app:12}, the above
discussion is only valid when the cross-sections are mainly independent of
the collision speeds $u_{i}^{cx}$. This is in general not true, even
for the reaction $H+p$ this holds only for energies (speeds) below
5\,keV ($\approx 1000$\,km/s) and thus the assumption of a nearly
constant cross-section may not be valid for high speed streams of the
solar wind, and not even for astrospheres of hot stars, where the
speed is of the order of a few 1000\,km/s. Thus higher order
approximations of the Taylor expansions are required. Because this
leads to clumsy expressions, which also need a lot of computational
effort, it may be better for practical purposes to solve the collision
integrals numerically as in \citet{Fichtner-etal-1996b}.

\section{Astrospheres revisited}\label{sheath}

Most of the aspects discussed above hold in general for
astrospheres. Nevertheless, because the stellar winds and the
interstellar medium can be largely different to that of the
heliosphere care must be taken which of the collision channels
displayed in Table~\ref{tab:1} and~\ref{tbb} needs to be taken into
account. For example, a stellar wind of 2000\,km/s has kinetic
energies around 20\,keV and most of the interactions become
important. Such wind speeds are derived by \citet{Vidotto-etal-2011}
and common in winds around hot stars
\citep{Arthur-2012,Decin-etal-2012}. For such astrospheres, we can
have strong bow shocks, because the interstellar wind is comparatively
strong. We may then encounter the situation where the low FIP (first
ionization potential) elements cannot reach the inner astrosheath, but
the high FIP elements do. This gives an interesting aspect when
discussing acceleration of energetic particles. Such a filtering of
low FIP elements happens in the heliosphere concerning carbon, which
becomes easily ionized already in the interstellar medium and thus
cannot penetrate into the heliosphere \citep{Cummings-etal-2002}.

In astrospheres additional effects can play a role, like elastic
collisions into an excited atomic state and subsequent photon
emission, thus energy loss. Recombination can take place, as well as
Bremsstrahlung or Synchrotron radiation, when a strong enough magnetic
field is present, which can lead to additional cooling of the stellar
wind plasma.

Astrospheres show a variety of shapes and all can be described by a
set of (hybrid) (M)HD equations discussed in Appendix~A.
If the surrounding interstellar medium is particularly ionized, it
  must be carefully checked which of the charge exchange reactions
  play a role. Especially, the charge exchange between particles with
  different masses needs to be carefully implemented, because in the
  collision integrals the reduced mass $m_{i}/(m_{i}+m_{j})$ (where
  $m_{i},m_{j}$ denote the corresponding mass of species $i,j$) appears
  as a factor. 

\section{Conclusions\label{sec:7} }

We have demonstrated that electron impact ionizations are important processes
inside the heliopause and particularly in the heliosheath, while other
charge exchange processes, except that of
H$^{+}$+H$\rightarrow$H$^{+}$+H, have been neglected because of their
small cross-sections for solar wind speeds. The latter become
important when modeling astrospheres, where the stellar wind speeds
can be larger by a factor 5 or more (up to 3000\,km/s
\citep{Arthur-2012}). In that case it is evident from Fig.~\ref{Fig2}
that a thorough discussion of the charge exchange between \Hy and \He
atoms, both neutral and ionized, is needed. The newly born PUIs heat
the stellar wind even inside the termination shock, so that the
threshold for electron impact ionization is exceeded, and here
electron impact plays a role, which then can further diminish the
neutral \He density by another 3\%.

Care must also be taken, because the cross-sections are taken from a
fit curve which may deviate by up to 10\% from the data, see also the
discussion by \citet{Bzowski-etal-2013}. Moreover, for the continuity,
momentum and energy equations the different collision speeds need to
be taken into the modeling efforts. Additionally, depending on the
representation of the cross-sections, the assumption that they are
nearly constant, as in \citet{McNutt-etal-1998,Heerikhuisen-etal-2008}
over a wide range, does not hold anymore. Therefore, either higher
order corrections in the Taylor expansion of the cross-sections need
to be taken into account, or a numerical estimation of the collision
integrals is needed in the modeling of the heliosphere or
astrospheres.

Especially for astrospheres, which can have relative speeds of a few
thousand kilometers per second, the non-resonant charge exchange
processes for \He become more and more important. For the huge
astrospheres of O-stars of tens of parsecs \citep{Arthur-2012} the
ionization of other species should be carefully considered, because
they can play a role in heating the supersonic stellar wind.

Also in the heliosphere during high-speed solar wind streams or
coronal mass ejections traveling into the outer regions of the
heliosphere the dynamics of the solar wind and propagating shock fronts
of CMEs will be influenced by charge exchange processes, including
electron impact, and by different \He ionizing processes.

For (M)HD simulations, one can use the governing equations, i.e.\ one
set of Euler equations for the combined ions and one for the sum of
neutrals. The densities of all included species can be handled with a
test particle approach, i.e.\ only solving the corresponding continuity
equation to get the correct density distribution in the heliosphere. A
better strategy is to solve in addition the energy equation for each
species to get their contribution to the total pressure
self-consistently. In the governing energy equation also a
sophisticated heat transport can be included. In a forthcoming work we
will include the above described processes in our model, especially add
a new species namely \He, as well as heat flow caused by the
electrons and study the relevance of all these effects.

Finally, there are clear indications that the electrons downstream of
the termination shock can easily attain much higher temperatures
compared to that of the protons \citep{Fahr-Chalov-2013}.

 We further point out that the He$^{+}$ abundance modeled by
  \citet{Slavin-Frisch-2008} reduces the Alfv\'en speed by about 10\%,
  which again shows the importance to include \He in models. 

We have shown here that all charge exchange processes needs
   to be reanalyzed carefully in order to minimize the errors in
  large-scale models and to improve the fits to the newly available IBEX
  and Voyager data.

\begin{acknowledgements}
  KS and HF are grateful to the
  \emph{Deut\-sche For\-schungs\-ge\-mein\-schaft, DFG\/} for funding
  the projects FI706/15-1 and SCHE334/10-1. MB was
  supported by the Polish Ministry for Science and Higher Education
  grant N-N203-513-038, managed by the National Science Centre.  SESF
  thanks the South African National Research Foundation for financial
  support.
\end{acknowledgements}

\appendix%
\section{Set of MHD equations\label{app:1} }

A general set of the Eulerian continuity-, momentum-, and energy equations
are the following \citep[see e.g.\ ][]{Boyd-Sanderson-2003}:
\begin{eqnarray}\label{ap:1}
\del \begin{bmatrix} \rho_{j} \\ \rho_{j} \vec{v}_{j} \\ E_{j}\\\vec{B} \end{bmatrix} 
+\div   \begin{bmatrix}
        \rho_{j} \vec{v}_{j} \\
        \rho_{j} \vec{v}_{j}\vec{v}_{j} + P_{j} \hat{I} - \dfrac{\vec{B}\vec{B}}{4\pi}\\[0.25cm]
        (E_{j} + P_{j})\vec{v}_{j}
        -\dfrac{\vec{B}(\vec{B}\cdot\vec{v_{j}})}{4\pi} \\
        \vec{v}_{j}\vec{B}-\vec{B}\vec{v}_{j}
         \end{bmatrix} =\\\nonumber
  \begin{bmatrix} 0 \\ \rho_{j} \vec{F} + \div \hat{\sigma}\\
      \rho_{j} \vec{v}_{j}\cdot\vec{F} + \div (\vec{v}_{j}\cdot\hat{\sigma}) -
      \div \vec{Q}\\0\end{bmatrix}
+ \begin{bmatrix} S_{j}^{c} \\ \vec{S}_{j}^{m}\\S_{j}^{e}\\A_{j}\end{bmatrix}
\end{eqnarray}

 \begin{tabular}{lll}
 $\vec{v}_{j}$&=& fluid velocity \\
 $\rho_{j}$ &=&  fluid density \\
 $E_{j}$ &=& internal energy of fluid \\
 $P_{j}$ &=& pressure of fluid  \\
$\hat{I}$ &=& unit tensor\\
$\hat{\sigma}$ &=& viscosity/stress tensor\\
$\vec{F}$ &=& external force per unit
 mass and volume\\
$\vec{Q}$ &=& heat flow\\
 $S_{i}^{j}$ &=& sources and sinks\\
$A_{j}$ &=& ambipolar diffusion between ions and neutrals
 \end{tabular}
~\\

Where the index $j\in\{H,H^{+},He,He^{+}, He^{++}, i, n\}$. As
discussed above for the index $j$ of the ionized fluid, the density is
the sum of all densities of ionized species, the total pressure
$P_{j}=\sum\limits_{k}P_{k}$,  the total energy
$E_{j}=\sum\limits_{k}E_{k}$, with $k\in\{$H$^{+},$He$^{+},$He$^{++}\}$. 

The force densities, stress tensor and the heat transport on the right hand
side of Eqs~\ref{ap:1} are usually neglected in modeling the large
scale of structure of the heliosphere, i.e.\
$\vec{F}=\hat{\sigma}=\vec{Q}=0$. While external forces, like the
solar gravitation, can be neglected, the stress tensor $\hat{\sigma}$
will play an important role in studying the details of the termination
shock or heliopause structure, but, to our knowledge was not
discussed so far. Especially the heat transport by electrons can
be expected to be significant, because of the high thermal speed of
these particles, and will be analyzed in future work. If the right
hand side of Eq.~\ref{ap:1} vanishes the set of equations is called
ideal MHD.

 From Eq.~\ref{eq:10} it is evident that the interstellar \He contributes about
  40\% to the total mass density, thus the total mass density $\rho_{n}$ in the above neutral
  continuity equation is $\rho_{n}=\rho_{H}+\rho_{He}\approx 1.4
  \rho_{H}$. This results in an increased ram pressure and momentum
  flow $\rho_{n}\vec{v_{n}}$. Together with the total pressure
  $P_{n}=P_{H}+P_{He} = \kappa (n_{H}T_{H}+n_{He}T_{He})$ and the
  previous made assumption $T_{H}=T_{He}$ and $n_{He}=0.1n_{H}$ yields
  $P_{n}=1.1 P_{H}$. The above estimates show, that \He contributions
  are not negligible. A similar consideration holds for the governing
  equations of the charged particles. 

If the \He inflow velocity differs in direction from the \Hy
\citep{Lallement-etal-2005}, then \He has to be treated as separate
fluid and an additional complete set of the above Euler-equations must be
solved. If one can assume that the flow velocities of the neutrals and
charged fluids  are the same for all neutral and ionized species, then
it is sufficient to solve the two governing equations and treat the
other species as tracer particles to calculate their densities and
thermal pressures. 

Handling the neutrals with a kinetic set of equations
\citep{Izmodenov-2007, Heerikhuisen-etal-2008} requires a similar
approach for the collision integrals (see Eq.~\ref{eq:i1} and~\ref{eq:i12}) to obtain
the balance terms $S^{c}_{j},\vec{S}^{m}_{j},S^{e}_{j}$ for the above
(M)HD equations of charged particles, when including heavier ions.

\section{``Collision'' speeds}\label{app:12}

For the interaction of particles with different masses (non-resonant
charge exchange processes), the approach by
\citet{McNutt-etal-1998} can be applied. Nevertheless, the relative velocities
for the momentum exchange differ from those for the energy exchange
in contrast to what is discussed in \citet{McNutt-etal-1998}:

\begin{figure}
  \includegraphics[width=0.95\columnwidth]{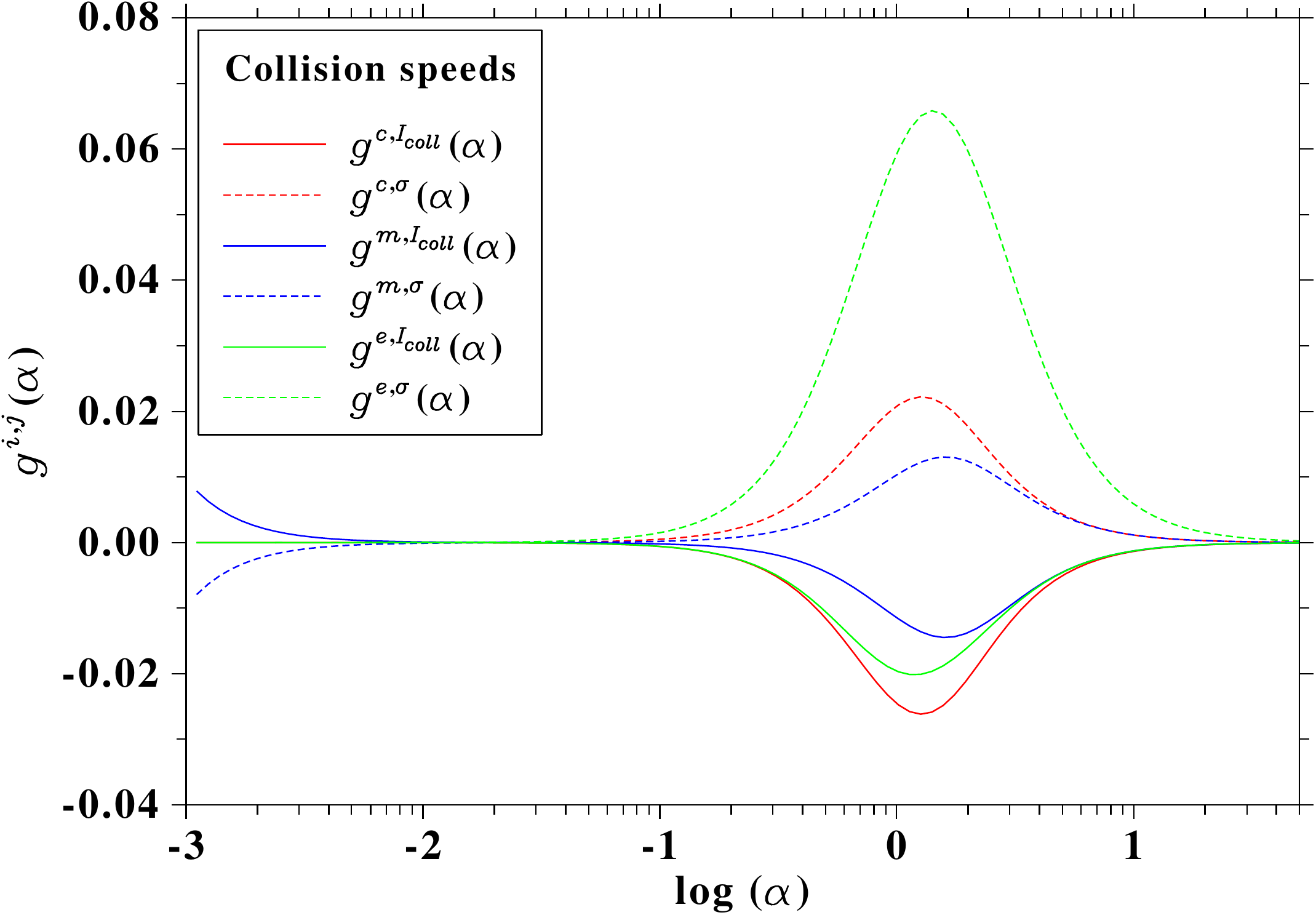}
  \caption{The relative errors
    $g^{i,j}=(u_{rel}^{i,j}-\tilde{u}_{rel}^{i,j})/u_{rel}^{i,j}$ of the fitted
    characteristic collision speeds, where $\tilde{u}_{j}^{i}$ are the
    approximations given in Eqs~\ref{eq:a2}-~\ref{eq:a7}.
  }\label{Figa2}
\end{figure}

\begin{eqnarray}\nonumber
  \label{eq:a1}
  w_{i}^{2} = \frac{k_{B}T_{i}}{2 m_{i}}; \qquad \kappa =
  \frac{1}{w_{1}^{2}+w_{2}^{2}}; \qquad \Delta \vec{v} = \vec{v}_{2} -
  \vec{v}_{1}; \\\nonumber
\Delta v = \sqrt{(\vec{v}_{2} -  \vec{v}_{1})^{2}}; \quad \alpha =
\Delta v \sqrt{\kappa}
\end{eqnarray}
As described in the formulas~57 to~60 in \citet{McNutt-etal-1998} the
relative speeds for the cross-sections for charge exchange differ for
all three Euler equations, as well as the speeds needed for the
charge exchange cross-section. The speeds are denoted as above
$u_{i}^{j}$, where $j\in\{c,m,e,P\}$ for the continuity, momentum and
energy equations and $j=P$ for the thermal pressure, respectively. 
The governing energy equation contains the ram pressure as well as the
thermal pressure, which are both influenced. The balance equations
only contain the thermal pressure, because changes in the ram pressure
or bulk speed are calculated in the governing equations due to the
assumption that all species flow with the same speed. The
index $i\in\{cx,rel\}$ are the speeds needed for the charge exchange
cross-section and corresponding relative speeds. With the definitions
from \citet{McNutt-etal-1998} we have to solve the following type of
integrals:
\begin{eqnarray}
  \label{eq:a12}
  I_{n}^{c,s} = \int\limits_{0}^{\infty}x^{n}e^{-\beta x^{2}}
  \left\{ \begin{array}{c}\sinh(\gamma
  x)\\ \cosh(\gamma x) \end{array}\right\} dx 
\end{eqnarray}
The following recursion holds:
\begin{eqnarray}
  \label{eq:a13}
  I_{n+2}^{c,s} = - \frac{\partial I^{c,s}_{n}}{\partial \beta}
\end{eqnarray}
Thus we need only to know the four integrals \citep{Gradstein-Ryshik-1981},
Nr. 3.562.3-3562.6:
\begin{eqnarray}\nonumber
  I_{1}^{c} &=& 
  \frac{\gamma}{4\beta}\sqrt{\frac{\pi}{\beta}}\exp{\left(\dfrac{\gamma^{2}}{4\beta}\right)}
    \erf\left(\frac{\gamma}{2\sqrt{\beta}}\right) + \frac{1}{2\beta} \\\nonumber
  I_{2}^{c} &=& \frac{\sqrt{\pi} (2\beta + \gamma^{2})}{8\beta^{2}
    \sqrt{\beta}}
  \exp{\left(\dfrac{\gamma^{2}}{4\beta}\right)}\\\nonumber
I_{1}^{s} &=& \frac{\gamma}{4\beta}
\sqrt{\frac{\pi}{\beta}}\exp{\left(\dfrac{\gamma^{2}}{4\beta}\right)}\\\nonumber
I_{2}^{s} &=&
\frac{\sqrt{\pi}(2\beta+\gamma^{2})}{8\beta^{2}\sqrt{\beta}}\exp{\left(\dfrac{\gamma^{2}}{4\beta}\right)}   \erf\left(\frac{\gamma}{2\sqrt{\beta}}\right) + \frac{\gamma}{4\beta^{2}}
\end{eqnarray}
The integrals above need to be multiplied by the factors given in 
\citet{McNutt-etal-1998}.

With these integrals it is easy to calculate the required speeds (for
details see \citet{McNutt-etal-1998}):
\begin{flalign}
  \label{eq:a2}
  u_{rel}^{c,I_{coll}} &= \frac{1}{\sqrt{\kappa}}
  \left(\frac{1}{2\alpha+\alpha}\right) \erf(\alpha)+
  \dfrac{e^{-\alpha^{2}}}{\sqrt{\pi}} \approx \frac{1}{\sqrt{\kappa}}\sqrt{\frac{4}{\pi}+\alpha^{2}}\\\label{eq:a3}
 u_{rel}^{c,\sigma} &= \frac{1}{\sqrt{\kappa}}
 \dfrac{\frac{3}{2}+\alpha^{2}}{u_{rel}^{c,I_{coll}}} \approx
 \frac{1}{\sqrt{\kappa}}\sqrt{\frac{9\pi}{16}+\alpha^{2}}\\\label{eq:a4}
u_{rel}^{m,I_{coll}} &= \frac{1}{\sqrt{\kappa}}
\left(-\frac{1}{4\alpha^{3}}+\frac{1}{\alpha}+\alpha
\right)\erf(\alpha) +  \left(2 +
  \frac{1}{\alpha^{2}}\right)\dfrac{e^{-a^{2}}}{\sqrt{\pi}}\\\nonumber
&\hspace{1cm}\approx\frac{2}{\sqrt{\kappa}}\sqrt{\frac{64}{9\pi}+\alpha^{2}}\\\label{eq:a5}
u_{rel}^{m,\sigma}&=
\frac{1}{\kappa}\dfrac{\frac{5}{2}+\alpha^{2}}{u_{rel}^{m,I_{coll}}}\approx \frac{1}{2\sqrt{\kappa}}\sqrt{\frac{225\pi}{256}+\alpha^{2}}\\\label{eq:a6}
u_{rel}^{e,I_{coll}}&=  \frac{1}{\sqrt{\kappa}} \left(\frac{3}{4\alpha} +
  3\alpha+\alpha^{3} \right) \erf{\alpha} +
\left(\frac{5}{2}+\alpha^{2} \right)\dfrac{e^{-\alpha^{2}}}{\sqrt{\pi}} \\\label{eq:a7}
u_{rel}^{e,\sigma} &= \frac{1}{\kappa}
\dfrac{\frac{15}{4}+5\alpha^{2}+\alpha^{4}}{2
  u_{rel}^{c,I_{coll}}+\alpha^{2}u_{rel}^{m,I_{coll}}}
\end{flalign}
The approximations in formulas~\ref{eq:a2} to~\ref{eq:a7} fit nicely,
except for small errors (see Fig~\ref{Figa2}). Note: The fit in
Eq.~\ref{eq:a5} differs by a factor 0.5 from that of
\citet{McNutt-etal-1998}. 

With the help of the integrals $I_{n}^{c,s}$ it is easy to calculate
the higher-order correction terms. It turns out that this is not
necessary for the case of the $p+H$ reaction, because the second-order
derivatives of the cross-sections, as shown in Fig.~\ref{Figa2} are orders of magnitudes
less then the cross-section. Shown in  Fig.~\ref{Figa3} are the
functional dependencies given by \citet{Fite-etal-1962} and by
\citet{Lindsay-Stebbings-2005}, i.e.
\begin{eqnarray}
  \sigma_{F} &=& [2.6\cdot10^{-7}-9.2\cdot10^{-9}\ln(v_{rel})]^{2}\\\nonumber
  \sigma_{LS}&=& [22.8-1.09\ln(v_{rel})]^{2}\cdot \left(1-\exp\left\{-\frac{1.3\cdot10^{17}}{v_{rel}^{2}}\right\}\right)^{2}
\end{eqnarray}
where $v_{rel}$ is given in cm/s and the $\sigma$ in cm$^{2}$. 

With the help of the computer algebra system wxMaxima
(\mbox{http://sourceforge.net/projects/wxmaxima/}) it is easy to
calculate the second derivatives of the above commonly used cross-sections
for the $p+H$ reaction and the integrals for the second order terms of
the Taylor-expansion for $\sigma^{cx}$. For the terms needed in the
continuity equation, we have to compare $\sigma^{cx}(u_{0})I^{s}_{2}$ with the
second order terms
$(I^{s}_{4}-2I^{s}_{3}v_{0}+I_{2}^{s}v_{0}^{2})\left.\frac{ \partial^{2}
\sigma}{ \partial v^{2}}\right|_{v_{0}}$, where we have for simplicity neglected
common factors. In our notation $v_{0} = I_{3}^{s}/I_{2}^{s}$. With
some algebra, we find that $I_{2}^{s}v_{0}^{2} >
I_{4}^{s}-2I^{s}_{3}v_{0}$ and thus finally:
\begin{eqnarray}
  \label{eq:aa1}
  h(v_{0}) \equiv \frac{v_{0}^{2}}{\sigma^{cx}(v_{0})}
  \left. \frac{\partial^{2} \sigma^{cx}(v)}{\partial v^{2}}\right|_{v_{0}}
\end{eqnarray}
The function $h(v)$ defined above should be small for all $v$, so
that the assumption can be made that the higher order terms in the Taylor
expansion of $\sigma^{cx}$ vanish. In Fig.~\ref{Figa3} the function is
shown for the two above discussed cross-sections.
\begin{figure}
  \includegraphics[width=0.95\columnwidth]{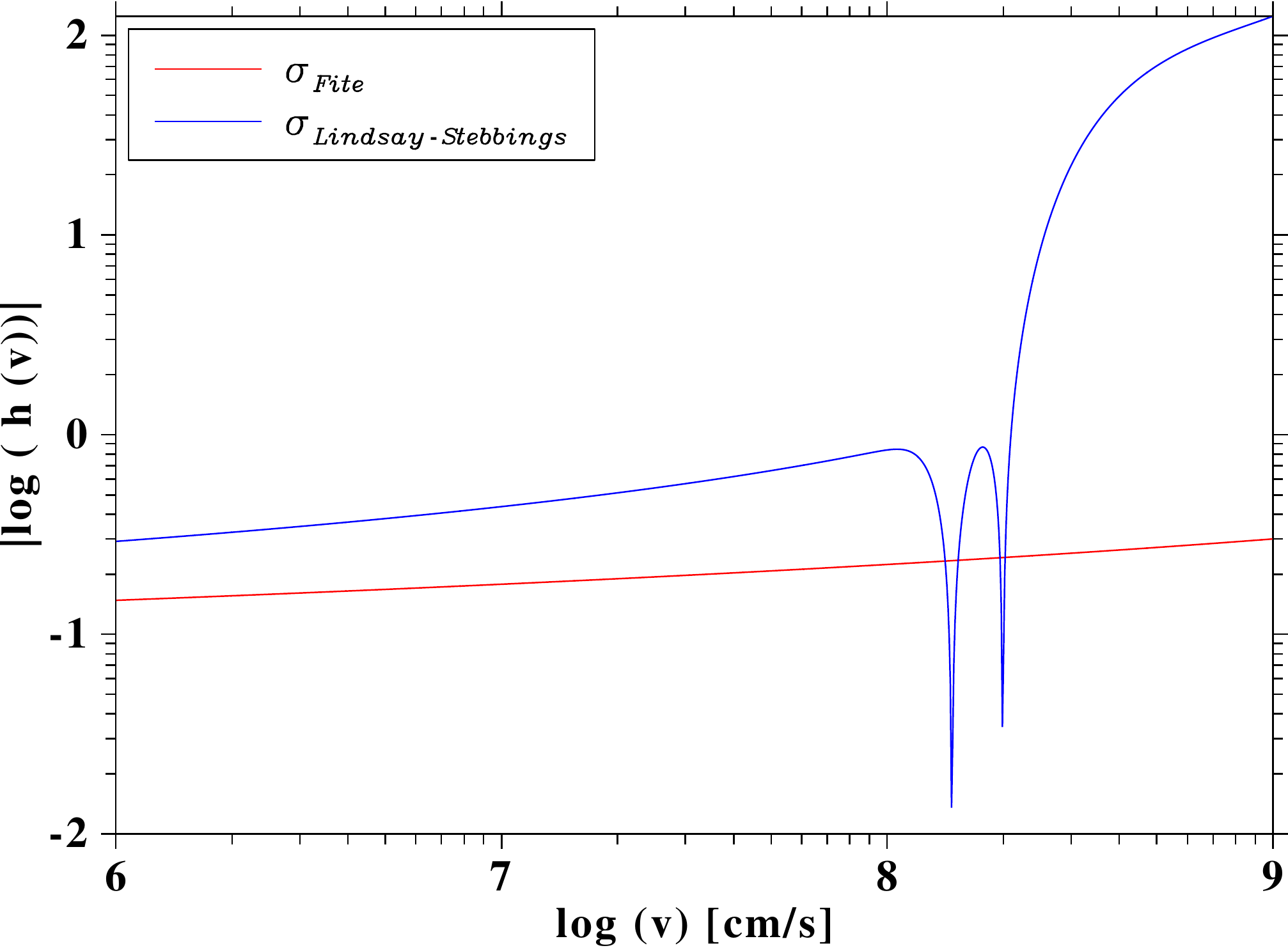}
  \caption{The ratio $h(v)$ (Eq.~\ref{eq:aa1}) of the second order
    integrated Taylor expansion to the zeroth order for both H-p
    charge exchange 
    cross-sections discussed by \citet{Fite-etal-1962} and
    \citet{Lindsay-Stebbings-2005}. The bump between the two downward
    spikes in the Lindsay-Stebbings curve is negative, but to be
    represented in a logarithmic scale its absolute value was
    taken.
    See text for restrictions of the fit functions. \label{Figa3}}
\end{figure}

It can be seen in Fig.~\ref{Figa3} that for the \citet{Fite-etal-1962}
cross-section the second order terms are nearly a factor 10 less than
the zeroth order term for all speeds. For the
\citet{Lindsay-Stebbings-2005} cross-sections this holds only for
speeds less than $4\cdot10^{7}$\,cm/s\,=\,400\,km/s. Around these speeds,
i.e.\ a commonly used solar wind speed, the second order corrections
are of the same order than the zeroth order term. 

In the latter case, i.e.\ if the higher order terms cannot be
neglected. In the analytic representation this will lead to clumsy
formulas, which also need some computational time. Thus, it might be
better to solve the collision integrals numerically.

Note,  for the hydrogen-proton reactions the fit function
  discussed in \citet{Fite-etal-1962} has an upper boundary of
  10\,keV, which corresponds to a relative speed of $\approx
  1400$km/s. The range in which the data are fitted by
  \citet{Lindsay-Stebbings-2005} has an upper boundary of 300 keV, which
  corresponds to a speed less than $\approx$7500\,km/s.
  The data presented in Fig.~\ref{Fig2} are taken from the
  ``Redbooks''
  (http://www-cfadc.phy.ornl.gov/redbooks/one/a/1a22.html) and range
  up to 63\,keV corresponding to $\approx 3400$km/s. Some of the \He
  reactions have their maximum cross-section at higher
  energies, as can be seen in Table~\ref{tab:1} and their upper limits
  in energy/amu are higher than shown in Fig.~\ref{Fig2}.  

All the fits given for the data from the ``Redbooks'' or the
  ``Aladdin'' database are mainly based upon a fit with Chebyshev
  polynomials. These polynomial fits are only good in the range given
  by the corresponding upper and lower (energy) boundaries and cannot
  extend beyond. Because usually in a large-scale model it cannot be
  guaranteed that always the relative speeds are in that range, a
  better fit function is needed, which can be extended beyond the
  mentioned boundaries keeping in mind not to violate quantum
  mechanical requirements.  

We have here only checked the simplest case, i.e.\ the collision terms
for the continuity equation. Of course such an analysis must not only be
done for the remaining set of MHD equations, but has, especially,
to be checked for all interactions.

\section{Interaction terms \label{app:2} }

In the following we give an overview of possible ionization
reactions. In table~\ref{tab:1}, the first column gives the interaction
term in the continuity equation, indicating the particle species:
X$^{n}$ with X$\in\{$H,He$\}$, while $n \in\{0,1,2\}$ gives the
ionization state (usually the index 0 is omitted). Newly created ions
are named as $PUI_{X^{n}}$ and energetic atoms by $ENA_{X}$. The
reaction rates have the indices $cx,ei,pi$ for charge exchange,
electron impact, and photo-ionization, respectively. The rates are
further denoted by $\nu(i,j)$ where $i$ describes different reactants,
while $j$ stands for different products of the same reaction $i$. The
tuple $(i,j)$ is given in the fourth column, where a minus sign stands
for losses and a plus sign for gains. To get the total gains and
losses one has to sum over all $i,j$.
In the fifth and sixth column the maximal cross-section and its
corresponding relative speed is given, respectively. In the third column
the first reactant is a particle from the solar wind, while the second
reactant is from the interstellar medium. This distinction is
necessary to ensure the correct sorting. 


Note: In the charge exchange rate $\nu^{cx} =
\rho_{X}\sigma^{cx}v_{rel}$ the density of the first reactant is
included, and indicated in the  table by $\nu_{cx,X}$.

Table~\ref{tab:1} contains the losses, while table~\ref{tbb} the
gains. In table~\ref{tab:1} the approximate maximal values for the
cross-section is given. It may vary by a factor of three and is only
supposed to give the correct order of magnitude. If for the
corresponding relative velocity a range is given, the cross-section is
a flat curve in that range.

\onecolumn
Table~\ref{tab:1}: The gains and losses from the original populations\\
\begin{table}
\caption{Losses}
\label{tab:1}
\begin{tabular}{llllrr}
                &
interactions &reactions&(i,j)&$\approx$max$(\sigma)$\,cm$^{2}$&$\approx v_{rel}$\,km/s\\
\hline
$S_{H^{+}}^{c}=$&$-\nu^{cx}_{H^{+}}\rho_{H}$&H$^{+}$+H$\rightarrow$H+H$^{+}$&-(1,1)&$10^{-15}$& $<400$\\
               &$-\nu^{cx}_{H^{+}}\rho_{He}$&H$^{+}$+He$\rightarrow$H+He$^{+}$&-(2,1)&$2\cdot10^{-16}$&$1100$\\
               &&H$^{+}$+He$\rightarrow$H$^{-}$+He$^{++}$&-(2,2)&$7\cdot10^{-18}$&$30$\\
               &$-\nu^{cx}_{H^{+}}\rho_{He^{+}}$&H$^{+}$+He$^{+}$$\rightarrow$H+He$^{++}$&-(3,1)&$2\cdot10^{-17}$&$1200$\\
               &$-\nu^{cx}_{H^{+}}\rho_{H^{-}}$& H$^{+}$+H$^{-}$$\rightarrow$H+H&-(4,1)&$10^{-14}$&$10-1000$\\
\hline
$S_{H^{-}}^{c}=$& $-\nu^{cx}_{H^{+}}\rho_{H^{-}}$& H$^{+}$+H$^{-}$$\rightarrow$H+H&-(4,1)&$10^{-14}$&$10-1000$\\
               &$-\nu^{cx}_{He^{+}}\rho_{H^{-}}$&He$^{+}$+H$^{-}$$\rightarrow$He+H& -(8,1)&$10^{-14}$&$10-1000$\\
\hline
$S_{He^{+}}^{c}=$&$ -\nu^{pi} \rho_{He^{+}}$&He$^{+}$+h$\nu\rightarrow$ He$^{++}+e$&-(5,1)& \\
               &$ -\nu^{ei} \rho_{He^{+}}$& He$^{+}$+e$\rightarrow$ He$^{++}$+2e &-(6,1)&$2\cdot10^{-17}$&$40000$\\
               &$-\nu^{cx}_{He^{+}}\rho_{H}$&He$^{+}$+H$\rightarrow$He+H$^{+}$&-(7,1)&$3\cdot10^{-16}$&$2400$\\
               &$-\nu^{cx}_{He^{+}}\rho_{H^{-}}$&He$^{+}$+H$^{-}$$\rightarrow$He+H& -(8,1)&$10^{-14}$&$10-1000$\\
               &$-\nu^{cx}_{He^{+}}\rho_{He}$&He$^{+}$+He$\rightarrow$He+He$^{+}$&-(9,1)&$10^{-15}$&$10-1000$\\
               &$-2\nu^{cx}_{He^{+}}\rho_{He^{+}}$&He$^{+}$+He$^{+}\rightarrow$He+He$^{++}$&-(10,1))&$2\cdot10^{-16}$&$6300$\\ 
               &$+\nu^{cx}_{He^{++}}\rho_{He}$&He$^{++}$+He$\rightarrow$He$^{+}$+He$^{+}$&+(11,2)&$3\cdot10^{-16}$&$1000$\\
               &$-\nu^{cx}_{He^{++}}\rho_{He^{+}}$&He$^{++}$+He$^{+}\rightarrow$He$^{+}$+He$^{++}$&-(12,1)$^{*}$&$5\cdot10^{-16}$&$100-1000$\\ 
\hline
$S_{He^{++}}^{c}=$&$-\nu^{cx}_{He^{++}}\rho_{He}$&He$^{++}$+He$\rightarrow$He+He$^{++}$&-(11,1)&$10^{-15}$&$1$\\
               &$ +\nu^{pi} \rho_{He^{+}}$&He$^{+}$+h$\nu\rightarrow$ He$^{++}+e$&+(5,1) \\
               &$+\nu^{cx}_{He^{+}}\rho_{He^{+}}$&He$^{+}$+He$^{+}$$\rightarrow$He+He$^{++}$&+(10,1)&$2\cdot10^{-16}$&$6300$\\ 
               &$ +\nu^{ei} \rho_{He^{+}}$& He$^{+}$+e$\rightarrow$ He$^{++}$+2e &+(6,1)&$2\cdot10^{-17}$&$100$\\
              &$-\nu^{cx}_{He^{++}}\rho_{He}$&He$^{++}$+He$\rightarrow$He$^{+}$+He$^{+}$&-(11,2)&$3\cdot10^{-16}$&$1000$\\
               &$-\nu^{cx}_{He^{++}}\rho_{He^{+}}$&He$^{++}$+He$^{+}$$\rightarrow$He$^{+}$+He$^{+}$&-(12,1)$^{*}$&$5\cdot10^{-16}$&$100-1000$\\ 
               &$-\nu^{cx}_{He^{++}}\rho_{H}$&He$^{++}$+H$\rightarrow$He$^{+}$+H$^{+}$&-(13,1)$^{*}$&$10^{-15}$&$2000$\\
\hline
$S_{H}^{c}$ =  &$ -\nu^{pi} \rho_{H}$&H+h$\nu\rightarrow$ H$^{+}+e$ &-(14,1)\\
               &$ -\nu^{ei} \rho_{H}$& H+e$\rightarrow$ H$^{+}$+2e &-(15,1)&$6\cdot10^{-17}$&$30000$\\
               &$-\nu^{cx}_{H^{+}}\rho_{H}$& H$^{+}$+H$\rightarrow$H$^{+}$+H$^{+}$+e&-(16,1)&$10^{-16}$&$2000$\\
               &&H$^{+}$+H$\rightarrow$H+H$^{+}$&-(1,1)&$10^{-15}$& $<400$\\
               &$-\nu^{cx}_{He^{+}}\rho_{H}$&He$^{+}$+H$\rightarrow$H$^{+}$+He&-(7,1)&$3\cdot10^{-16}$&$2400$\\
               &$-\nu^{cx}_{He^{++}}\rho_{H}$&He$^{++}$+H$\rightarrow$He$^{+}$+H$^{+}$&-(13,1)&$10^{-15}$&$2000$\\
               &$+\nu^{cx}_{H^{+}}\rho_{H^{-}}$& H$^{+}$+H$^{-}$$\rightarrow$H+H&+(4,1)&$10^{-14}$&$10-1000$\\
               &$+\nu^{cx}_{He^{+}}\rho_{H^{-}}$&He$^{+}$+H$^{-}$$\rightarrow$He+H& +(8,1)&$10^{-14}$&$10-1000$\\
\hline
$S_{He}^{c}$ = &$ -\nu^{pi} \rho_{He}$&He+h$\nu\rightarrow$ He$^{+}+e$&-(17,1)\\
               &$ -\nu^{pi} \rho_{He}$&He+h$\nu\rightarrow$ He$^{++}+2e$ &-(17,2)\\
               &$ -\nu^{ei} \rho_{He}$& He+e$\rightarrow$ He$^{+}$+2e &-(18,1)&$3\cdot10^{-17}$&$30000$\\
               &$-\nu^{cx}_{H^{+}}\rho_{He}$&H$^{+}$+He$\rightarrow$H+He$^{+}$&-(2,1)&$2\cdot10^{-16}$&$1100$\\
               &&H$^{+}$+He$\rightarrow$H$^{-}$+He$^{++}$&-(2,2)&$7\cdot10^{-18}$&$30$\\
               &$-\nu^{cx}_{He^{+}}\rho_{He}$&He$^{+}$+He$\rightarrow$He+He$^{+}$&-(9,1)&$10^{-15}$&$10-1000$\\
               &$-\nu^{cx}_{He^{++}}\rho_{He}$&He$^{++}$+He$\rightarrow$He+He$^{++}$&-(11,1)&$10^{-15}$&$1$\\
              &&He$^{++}$+He$\rightarrow$He$^{+}$+He$^{+}$&-(11,2)&$3\cdot10^{-16}$&$1000$\\
\hline
\end{tabular}
\end{table}

\begin{multicols}{2}
  All reactions that have as the second reactant
  a charged particle cannot flow in from the interstellar medium.
  Because the heliopause is a contact discontinuity separating 
   the solar wind and interstellar
  medium plasma, i.e.\ there is no flow of charged particles across
  the heliopause. Thus in the tables~\ref{tab:1} and~\ref{tbb} the reactions
  labeled (3,1), (4,1), (5,1), (6,1), (8,1), (10,1), and (12,1) can be
  neglected. 

  The reactions labeled with $^{*}$ are sorted in pickup channels,
  even though they do not create new PUIs. For example the reaction
  He$^{++}$+H$\rightarrow$He$^{+}$+H$^{+}$ produces a new pickup
  \Hy, but the change from doubly- to singly-charged \He does
  not change its character, i.e.\ the He$^{+}$ has the same bulk speed
  as the He$^{2+}$. Only the charge is changed, thus the new He$^{+}$
  does not affect the dynamics, (the H$^{+}$ does). Nevertheless,
  these process can have an effect on magnetic field turbulence
  \citep[e.g.\ ][]{Shalchi-etal-2012}. This aspect is interesting in the description of the
  diffusion tensor to model cosmic rays
  \citep{Effenberger-etal-2012}. The role of these type of processes
  for the turbulence will also be considered in future work.

\end{multicols}

\newpage
Table~\ref{tbb}: The gains and losses from the new populations:\\
\begin{table}
\caption{Gains}
\label{tbb}
\begin{tabular}{llll}
 Interaction &                                &reactions&(i,j)\\
\hline
$S_{H^{+}}^{c}$=&$ +\nu^{pi} \rho_{H}$&H+h$\nu\rightarrow$ H$^{+}+e$ &+(14,1)\\
                    &$ +\nu^{ei} \rho_{H}$& H+e$\rightarrow$ H$^{+}$+2e &+(15,1)\\
                 &$+\nu^{cx}_{H^{+}}\rho_{H}$&H$^{+}$+H$\rightarrow$H+H$^{+}$&+(1,1)\\
                 && H$^{+}$+H$\rightarrow$H$^{+}$+H$^{+}$+e&+(16,1)\\
             &$+\nu^{cx}_{He^{+}}\rho_{H}$&He$^{+}$+H$\rightarrow$H$^{+}$+He&+(7,1)\\
             &$-\nu^{cx}_{He^{++}}\rho_{H}$&He$^{++}$+H$\rightarrow$He$^{+}$+H$^{+}$&+(13,1)\\
\hline
$S_{H^{-}}^{c}$=&$+\nu^{cx}_{H^{+}}\rho_{He}$&H$^{+}$+He$\rightarrow$H$^{-}$+He$^{++}$&+(2,2)\\
\hline
$S_{He^{+}}^{c}$=& $ +\nu^{pi} \rho_{He}$&He+h$\nu\rightarrow$ He$^{+}+e$&+(17,1) \\
               &$ +\nu^{ei} \rho_{He}$& He+e$\rightarrow$ He$^{+}$+2e &+(18,1)\\
               &$+\nu^{cx}_{H^{+}}\rho_{He}$&H$^{+}$+He$\rightarrow$H+He$^{+}$&+(2,1)\\
              &$+\nu^{cx}_{He^{+}}\rho_{He}$&He$^{+}$+He$\rightarrow$He+He$^{+}$&+(9,1)\\
              &$+\nu^{cx}_{He^{++}}\rho_{He}$&He$^{++}$+He$\rightarrow$He$^{+}$+He$^{+}$&+(11,2)\\
               &$+\nu^{cx}_{He^{++}}\rho_{H}$&He$^{++}$+H$\rightarrow$He$^{+}$+H$^{+}$&+(13,1)$^{*}$\\
\hline
$S_{He^{++}}^{c}$=&$ -\nu^{pi} \rho_{He}$&He+h$\nu\rightarrow$ He$^{++}+2e$ &+(17,2)\\
                &&H$^{+}$+He$\rightarrow$H$^{-}$+He$^{++}$&+(2,2)\\
              &$+\nu^{cx}_{He^{+}}\rho_{He^{+}}$&He$^{+}$+He$^{+}$$\rightarrow$He+He$^{++}$&+(10,1)\\ 
              &$+\nu^{cx}_{He^{++}}\rho_{He}$&He$^{++}$+He$\rightarrow$He+He$^{++}$&+(11,1)\\
              &&He$^{++}$+He$^{+}$$\rightarrow$He$^{+}$+He$^{++}$&+(12,1)$^{*}$\\
\hline
$S_{H^{0}}^{c}$ = &$+\nu^{cx}_{H^{+}}\rho_{H}$&H$^{+}$+H$\rightarrow$H+H$^{+}$&+(1,1)\\
               &$+\nu^{cx}_{H^{+}}\rho_{He}$&H$^{+}$+He$\rightarrow$H+He$^{+}$&+(2,1)\\
               &$+\nu^{cx}_{H^{+}}\rho_{He^{+}}$&H$^{+}$+He$^{+}$$\rightarrow$H+He$^{++}$&+(3,1)\\
               &$+2\nu^{cx}_{H^{+}}\rho_{H^{-}}$& H$^{+}$+H$^{-}$$\rightarrow$H+H&2(+(4,1))\\
              &$+\nu^{cx}_{He^{+}}\rho_{H^{-}}$&He$^{+}$+H$^{-}$$\rightarrow$H+He& +(8,1)\\
\hline
$S_{He^{+}}^{c}$ = &$+\nu^{cx}_{He^{+}}\rho_{H}$&He$^{+}$+H$\rightarrow$H$^{+}$+He&+(7,1)\\
              &$+\nu^{cx}_{He^{+}}\rho_{H^{-}}$&He$^{+}$+H$^{-}$$\rightarrow$H+He& +(8,1)\\
              &$+\nu^{cx}_{He^{+}}\rho_{He}$&He$^{+}$+He$\rightarrow$He+He$^{+}$&+(9,1)\\
              &$+\nu^{cx}_{He^{+}}\rho_{He^{+}}$&He$^{+}$+He$^{+}$$\rightarrow$He+He$^{++}$&+(10,1)\\ 
              &$+\nu^{cx}_{He^{++}}\rho_{He}$&He$^{++}$+He$\rightarrow$He+He$^{++}$&+(11,1)\\
              &$+\nu^{cx}_{He^{+}}\rho_{He}$&He$^{+}$+He$\rightarrow$He+He$^{+}$&+(9,1)\\
              &$+\nu^{cx}_{He^{++}}\rho_{He}$&He$^{++}$+He$\rightarrow$He+He$^{++}$&(11,1)\\
\end{tabular}
\end{table}

\begin{multicols}{2}

From the above tables the interaction terms for the governing equations
can be determined: For the neutral equation it is just the sum of all
$S_{H}^{c}+S_{He}^{c}$ interaction terms. More care must be taken for
the ion governing equation, because the pickup terms can cancel some
of the ion terms, for example the reaction
H$^{+}$+H$\rightarrow$H+H$^{+}$ looses an original fast solar wind proton,
but gains a slow interstellar \Hy atom, which does not change the
total ion mass.

To describe the interaction terms for the momentum
and energy equation,  similar tables can now easily be constructed
and the governing equations determined.

\end{multicols}
\end{document}